\documentclass{statsoc}

\usepackage[a4paper]{geometry}
\usepackage{graphicx}
\usepackage[textwidth=8em,textsize=small]{todonotes}
\usepackage{amsmath}
\usepackage{natbib, url}

\usepackage{paralist}
\usepackage{hyperref}
\usepackage{mathtools}
\usepackage{amssymb}

\numberwithin{equation}{section}

\renewenvironment{quote}%
               {\list{}{}%
                \item[]}%
               {\endlist}

\newcommand{\bi}{\begin{itemize}}
\newcommand{\ei}{\end{itemize}}
\newcommand{\be}{\begin{enumerate}}
\newcommand{\ee}{\end{enumerate}}
\newcommand{\bq}{\begin{quote}}
\newcommand{\eq}{\end{quote}}

\newlength{\cslhangindent}
\newlength{\csllabelwidth}
\newlength{\cslentryspacingunit} 
 {
  \setlength{\parindent}{0pt}
  \ifodd #1
  \let\oldpar\par
  \def\par{\hangindent=\cslhangindent\oldpar}
  \fi
  \setlength{\parskip}{#2\cslentryspacingunit}
 }%

 \expandafter\def\expandafter\quote\expandafter{\quote\singlespacing}  

\def\bfseries{\fontseries \bfdefault \selectfont \boldmath}

\title[A Partially Pooled NSUM Model]{A Partially Pooled NSUM Model: Detailed estimation of CSEM trafficking prevalence in Philippine municipalities}
\author[Nyarko-Agyei {\it et al.}]{Albert Nyarko-Agyei$^1$*, Scott Moser$^2$, Rowland G Seymour$^3$, Ben Brewster$^4$, Sabrina Li$^6$, Esther Weir$^1$,Todd Landman$^1$, Emily Wyman$^1$, Christine Belle Torres$^5$, Imogen Fell$^5$, Doreen Boyd$^6$}
\address{$^1$Rights Lab, University of Nottingham, Nottingham, UK}
\address{$^2$Politics and International Relations, University of Nottingham, UK}
\address{$^3$School of Mathematics, University of Birmingham, UK}
\address{$^4$Nottingham University Business School, University of Nottingham, UK}
\address{$^5$International Justice Mission Philippines, Manila, Philippines}
\address{$^6$School of Geography, University of Nottingham, UK}
\email{albert.nyarko-agyei@nottingham.ac.uk*}

\begin{document}
\begin{abstract}
Effective policy and intervention strategies to combat human trafficking for child sexual exploitation material (CSEM) production require accurate prevalence estimates. Traditional Network Scale Up Method (NSUM) models often necessitate standalone surveys for each geographic region, escalating costs and complexity. This study introduces a partially pooled NSUM model, using a hierarchical Bayesian framework that efficiently aggregates and utilizes data across multiple regions without increasing sample sizes. We developed this model for a novel national survey dataset from the Philippines and we demonstrate its ability to produce detailed municipal-level prevalence estimates of trafficking for CSEM production. Our results not only underscore the model’s precision in estimating hidden populations but also highlight its potential for broader application in other areas of social science and public health research, offering significant implications for resource allocation and intervention planning.
\end{abstract}

\section{Introduction}
The internet has had a transformative impact on both consensual and non-consensual sex markets. It reduces the exposure of illict producers and buyers to police efforts to disrupt them and provides fora where sexual transactions can be discussed and facilitated, and payments made \citep{Perer12, Chan2019}. One area where this impact has been particularly profound is in relation to the production and dissemination of child sexual exploitation and abuse material. Moreover there is a range of different modes of exploitation with recent trends pointing towards the exploitation of children by co-located adult facilitators who are financially compensated by overseas sexual offenders in exchange for livestreamed sexual exploitation - a specific from of Child Sexual Exploitation Material (CSEM) (IJM, 2020).

The issue is emphasised in low- and middle-income countries, where there are limited resources to prevent and support children being trafficked by co-located adults to generate CSEM \citep{ECPAT_2022}. When tackling trafficking to produce CSEM at a national scale, it is necessary to understand where children are being abused so that authorities can identify where resources need to be allocated. 

The focus on human trafficking has grown significantly over the past two decades, with international efforts to combat its different forms—such as the online sexual exploitation of children—finding common ground under the term ``modern slavery" \citep{Bales12, Landman2020}. The push to measure these activities has also gained significant traction, serving both as indicators of the progress in combating these issues and as tools to rally support from governments and key stakeholders seeking to provide programmes of intervention.

As a largely `hidden' issue, measuring the extent of online sexual exploitation of children can require varied approaches because of the effect of several factors. These include cultural attitudes, the physical setting of the exploitation, political commitment, and other socio-economic considerations, as well as the mode of exploitation and the \textit{modus operandi} of those involved. For example, cultural and community norms can influence people's willingness to report or disclose information related to criminal activity. Additionally, the quality, structure, and methods used by state and non-governmental organizations to collect data and information affect how much is known about a specific issue. Moreover, the degree of awareness among key stakeholders, such as the private sector, government agencies, and civil society, can play a significant role in mobilizing organizations to address the problem.

The Network Scale Up Method (NSUM) estimates the size of hidden groups through Aggregated Relational Data (ARD) from surveys \citep{Bernard_1991, Killworth_1998b}. ARD is typically a matrix of the number of people known by each respondent from each group. A simple explanation of NSUM models applied to ARD can be broken down into two parts. In the first part, the size of each respondent’s network (degree) is estimated by asking them how many people they know in populations with known sizes (e.g. teachers or nurses). In the second part, respondents are asked how many members of the hidden group of interest they know. These two parts are then combined to estimate the proportion of each respondent’s network belonging to the hidden group. By aggregating across respondents an estimate of the size of the hidden group can be obtained.

NSUM models have been deployed in a wide range of applications, including to estimate the prevalence of homelessness \citep{McCormick_2012}, of abortion in Ethiopia and Uganda \citep{Sully_2020}, and of HIV in Singapore \citep{Quaye_2023}. The method is not without bias; the three main biases as described in \citet{Killworth_1998b} are: barrier bias, when groups are not randomly spread in the network of respondents so the random mixing assumption that respondent's networks are representative of the population is violated; transmission bias, when respondents do not have correct knowledge of the groups in their networks; and recall bias, when respondents can not accurately recall how many people are in each group. Variants of NSUM models that incorporate structure to account for these biases include a hierarchical Bayesian model proposed by \citet{Maltiel_2015} and collecting additional respondent level information to estimate bias by \citet{Laga_2023}. 

Current NSUM variants are, however, limited to applications involving only one geographical region and are not suited to estimating the size of hidden groups in multiple regions simultaneously. For example, running a national NSUM survey would provide a national prevalence estimate, but not provide rigorous estimates at sub-national level. To produce these sub-national estimates using current methods, an independent NSUM survey would have to be run in each sub-national region in the study. This would considerably increase the sample size compared to running a single national survey, incurring large financial costs and administrative burdens for facilitating the sub-national surveys. 

Our modelling aim is to develop a NSUM model that can provide estimates for the size of groups in multiple geographic regions without incurring a prohibitively large sample size. We seek to achieve this aim by developing a partially pooled variant of NSUM models. Using a hierarchical structure, we assume there is a national level distribution of group sizes from which the size of groups in each region is drawn. This assumption allows us to estimate the size of the hidden groups in each region simultaneously, without running standalone surveys in each region.

\subsection{Empirical motivation}
We are motivated by the Scale of Harm project, which set out to determine the prevalence of the trafficking of children in person to produce CSEM in the Philippines. This trafficking is mainly for foreign offenders who watch and direct abuse in real time for a fee and receive newly produced images and videos \citep{ijm_report_2023}. The Philippines is well established as a centre for online CSEM production. High levels of English language skills, availability of internet access, and high levels of deprivation mean there is a large pool of children who have the potential to be trafficked into producing exploitation material for dissemination to users in countries such as the USA, Australia, Germany, and UK \citep{DeMarco18, ECPAT_2022}. However, previous attempts to quantify the scale of the problem in the Philippines have not been able to produce national or regional estimates \citep{ijm_report_2020}. This is due in part to the difficulty in using existing CSEM material as indicative of current prevalence of production since many of these are instances of CSEM already in circulation rather than current cases of trafficking.

The production of online CSEM poses a serious danger to children and society, and children who are trafficked to produce such material are our motivating hidden group. We worked with International Justice Mission (IJM) Philippines and its Centre to End the Online Sexual Exploitation of Children. Professionals in IJM Philippines are carrying out long term interventions to enhance justice system response to CSEM. 

Our aim was to estimate the number of people in the Philippines who traffic children to produce CSEM, including an estimate for the prevalence in 150 of the 1,634 municipalities in the country. The need for an estimate at municipality level was to underpin the development of a national intervention programme and to inform the allocation of safeguarding and law enforcement resources by IJM.

To design the national household survey, we first carried out an eight-month period of workshops and interviews with health care, finance, and legal professionals, community leaders, and survivors of child sex trafficking in the Philippines. We organised this exploratory phase to decide the best approach to estimating the prevalence of trafficking to produce CSEM and the wording of survey questions, particularly sensitive questions about CSEM production. This period highlighted that information about the trafficking of children to produce CSEM is common knowledge in local communities in the Philippines, due to their close-knit nature and organisation into small administrative units known as barangays. We therefore developed a survey about online safety in the Philippines, including questions to collect ARD on trafficking to produce CSEM. Guidance on how to ask these questions and where they should be placed in the survey was provided by a group of survivors of trafficking to produce CSEM. Our questions probing the knowledge of the target hidden populations were:
\begin{enumerate}
  \item How many adults do you know that are involved in selling sexually explicit online livestreams of children?
  \item How many adults do you know that are involved in selling sexually explicit photos, OR videos online of children they know? 
  \item How many children do you know who, this year, have been used or forced by an adult to produce sexually explicit livestreams?
  \item How many children do you know who, this year, have been used or forced by an adult to produce sexually explicit photos or videos?
\end{enumerate}

Defining what it means to know someone in an NSUM survey needs to be carefully tailored to the community in which the survey is being run. We worked with survivors of online child sexual abuse, safeguarding professionals, and academics in the Philippines to determine a suitable definition of knowing someone. We defined `knowing' as someone in your municipality who a respondent recognised by both sight and name. We also stipulated for adult groups, that this had to likely be reciprocated, so to avoid reports of knowing a celebrity. This is a stringent definition, and we may have missed cases where respondents were aware of traffickers in their community but did not know their name.  Indeed, although community members are often able to say exactly which houses online child sexual abuse is happening , they may refer to traffickers with nicknames like `momshie' or `auntie' \citep{Ramiro2019}. We chose to include knowing by name in our definition to avoid respondents reporting hearsay or unfounded accusations about people they \textit{heard of} but did not know.

The survey was translated into five languages spoken in the Philippines and deployed nationally with a representative sample of 3,600 Filipino households taking part. This sample size was chosen as this corresponds to a 99\% confidence level and a 2.15\% margin of error with an average network size of 300 using the NSUM sample size heuristic described in \cite{Josephs2022}. This was chosen to satisfy the IJM's tolerance for uncertainty in the results. The survey was facilitated by IPSOS Philippines. Ethical approval was given by the University of Nottingham School of Sociology Ethics Committee and the Philippines Social Science Council Ethics Review Board. 

The remainder of the paper is structured as follows. In Section \ref{sec: model} we describe the standard NSUM model and our partially pooled version. In Section \ref{sec: sim study} we run simulation studies to evidence the ability of our model to provide municipality level estimates without an increased sample size. In Section \ref{sec: results} we fit our model to data we collected in the Philippines. In Section \ref{sec: discussion} we discuss our model results, limitations and future work. 

\section{Incorporating a pooled structure into the Network Scale Up Method} \label{sec: model}

\subsection{The Standard NSUM Model}
Consider a population of size $N$ that is categorized into $K$  potentially overlapping groups, where the size of group $k$ is denoted by $N_k$. We wish to infer the value of $N_k$ for groups that are hidden based on aggregated relational data collected in a survey. For the data, we denote the number of individuals in group $k$ that are reported known by respondent $i$ by $y_{ik}$ for $ 1 \leq k \leq K$ and $ 1 \leq i \leq R$. 

We now describe the standard NSUM model as proposed in \citet{Killworth_1998b}. The number of people in group $k$ that person $i$ knew is modelled by
$$
Y_{ik} \mid d_i \sim \hbox{Bin}\left(d_i, \frac{N_k}{N}\right),
$$
where $d_i$ is the number of people in person $i$'s personal network. Although the Binomial model is conditioned on $d_i$, in practice $d_i$ is not known and its maximum likelihood estimate given the groups of known size is used in place.  The authors assumed that the responses are independent across groups and individuals. From the standard NSUM model when $d_i$ is known, the maximum likelihood estimate for the size of a unknown group $k$ is 
$$\hat{N_k} = N\cdot\frac{\sum_{i=1}^R y_{ik}}{\sum_{i=1}^R d_i}.$$

In a study seeking to determine the size of a hidden group in multiple municipalities, using this standard NSUM model would treat each municipality's ARD separately with no information shared across the municipalities to generate estimates. This is plausible in contexts where the municipalities are clearly unrelated but this is not the case in subnational studies like our motivating example. 

\subsection{The Partially Pooled Network Scale Up Model}
To allow for information about the size of the hidden population in one municipality to be shared to other municipalities, we now develop the partially pooled NSUM model. We assume the size of the hidden group in each municipality is drawn from a country-wide distribution. We show how this sharing of information between municipalities allows us to reduce the sample size, and improve estimation relative to the standard NSUM model. 

\subsubsection{Modelling aggregated relational data}
 In the partially pooled NSUM model, we assume that there are $M$ municipalities and each respondent $i$ reports the number of people they know in group $k$ in municipality $m$, denoted $y_{ikm}$. Our aim is to estimate the number of people in the hidden group $k$ in municipality $m$, $N_{km}$, where the total population size of $m$, $N_m$, is known. 
 To account for overdispersion in the data, we follow \citet{Zheng_2006} and use a Negative Binomial distribution to model the responses from each survey respondent and assume
 \begin{equation}
    Y_{ikm} \sim \hbox{NegBin}\left(\exp(\delta_i - \rho_{km} \right), w_{k}).
\end{equation}
We define the expectation of this distribution through the link function $\exp(\delta_i - \rho_{km})$. The parameter $-\rho_{km}=\log\left(\frac{N_{km}}{N_m}\right)$ is the log of the proportion of the municipality $m$ in group $k$. The parameter $\delta_i$ represents the log of the number of people in respondent $i$'s social network $d_i$. The parameter $w_k > 0$ controls for the overdispersion in knowing people in group $k$, with 
$$
\hbox{Var}(Y_{ikm}) = \exp(\delta_i - \rho_{km})\left( 1+ \frac{\exp(\delta_i - \rho_{km})}{w_k}\right).
$$
As $w_k \to \infty$, the Negative Binomial distribution approaches a Poisson distribution with expectation $\exp(\delta_i - \rho_{km})$.  We parameterise this distribution through the overdispersion as this allows us to directly model a wide range of ARD that has a variety of overdispersion rates in the different groups. Hidden groups are typically marginalized or stigmatized populations so when examining ARD collected on hidden groups, many respondents will report no contact with the group in questions. However, some respondents will report having high levels of contact to the same groups. The Negative Binomial distribution allows us to model this phenomenon. The issue of differing levels of contact and dispersion has been raised in related NSUM literature, for example, when studying marginalized groups such as prisoners in \citet{Zheng_2006}. In the Supplementary Material, we provide a discussion of how the Negative Binomial model from \citet{Zheng_2006} arises and can be used to account for overdispersion in the data.

\subsubsection{Partial pooling}
We now develop a partially pooled framework using a hierarchical Bayesian framework, allowing us to draw upon both local and broader regional ARD to inform our prevalence estimates. We introduce the partial pooling through the parameter $\rho_{km}$, which is vital for pooling information across municipalities and thereby enhancing the reliability and accuracy of our prevalence estimates. The choice of the distribution for $\rho_{km}$ is a modelling choice. We choose the Gamma distribution, $\rho_{km} \sim \Gamma(\alpha_k, \beta_k)$. This provides us with a flexible shape to model a wide variety of group sizes across the municipalities. In particular, the shape of its tail allows us to capture situations where known groups, such as teachers, may have quite high prevalence, but hidden groups, such as children being trafficked for CSEM, are small, without placing considerable weight on these populations being unreasonably small. The shared expectation and variance for the number of people in each group $k$ is given by 
$$
\mu_{\rho_{k}} = \frac{\alpha_k}{\beta_k} \qquad \sigma^2_{\rho_{k}} = \frac{\alpha_k}{\beta_k^2}.
$$
To avoid confounding the model as to which parameter controls the size of the group and how much it is dispersed across the municipalities we place prior distributions directly on the mean and variance  $\mu_{\rho_{k}}$ and $\sigma^2_{\rho_{k}}$ respectively rather than on $\alpha_k$ and $\beta_k$. We place independent and identical, half-normal prior distributions on $\mu_{\rho_{k}}$ and $\sigma^2_{\rho_{k}}$ for all groups $k$ such that $\mu_{\rho_{k}} \sim HN(0, \sigma^2_{\mu_{\rho}})$ and $\sigma^2_{\rho_{k}} \sim HN(0, \tau^2)$. For the overdispersion parameter $w_k$, we use independent and identical half normal prior distributions for each group such that $w_k \sim HN(0, \sigma^2_w)$.  For the log of the number of people in respondent $i$'s network $\delta_i$, we follow \citet{Laga_2023} and assume $\delta_i \sim N(\mu_\delta, \sigma^2_\delta)$. 

We are motivated by recent extensions of Bayesian NSUM models, especially \citet{Laga_2023}.  While we try to stay as close to their model, in spirit, in this section we briefly discuss comparisons between the two approaches. First, \citet{Laga_2023} integrate additional respondent-level information to estimate biases, specifically focusing on correlated data within social networks to improve the estimation accuracy for hidden populations. They address transmission bias, which occurs when respondents lack accurate knowledge of group sizes within their networks, and barrier bias, which arises when hidden populations are unevenly distributed within respondents’ social networks.

While \citet{Laga_2023} incorporate extensive data to estimate these biases, our approach does not explicitly model transmission bias due to the impracticality of estimating it solely from ARD. Instead, our hierarchical Bayesian framework accounts for variability and overdispersion in the data. For example, estimating transmission bias directly in our study on trafficking in the Philippines would require detailed knowledge of each respondent's social network, which is often impractical.

Lastly, another critical difference lies in the use of priors. \citet{Laga_2023}'s formulation allows for unrealistic parameter estimates due to the priors used, potentially leading to invalid probabilities. Their prevalence parameter for group $k$ is given by $\exp(\rho_k) = \frac{N_k}{N}$.  As such $\rho_k$ support should be non-positive given that it represents a proportion, however a Normal distribution is used there. In contrast, our model uses a reflected Gamma distribution to model the proportions of hidden populations. 

\subsection{Implementing the model}
By Bayes' theorem, the posterior distribution is
$$
\pi(\mu_{\rho_{k}}, \sigma^2_{\rho_{k}}, \boldsymbol{\rho}, \boldsymbol{w}, \boldsymbol{\delta}, \sigma^2_\delta \mid \boldsymbol{y}) \propto \pi(\boldsymbol{y} \mid \boldsymbol{\rho}, \boldsymbol{w}, \boldsymbol{\delta}) \pi(\boldsymbol{\rho} \mid \mu_{\rho_{k}}, \sigma^2_{\rho_{k}}) \pi(\mu_{\rho_{k}})\pi(\sigma^2_{\rho_{k}})\pi(\boldsymbol{\delta} \mid \sigma^2_\delta)\pi(\sigma^2_\delta),
$$
where $\boldsymbol{\rho}$ represents the matrix of $\rho_{km}$, $\boldsymbol{w}$ is the vector of $w_k$ for each group, $\boldsymbol{\delta}$ is the vector of $\delta_i$ and $\boldsymbol{y}$ is the ARD. We developed a modelling fitting procedure in Stan and we used \texttt{R} as the interface to perform sampling. Running the model requires the total number of respondents across all municipalities, the number of groups, the ARD, the number of municipalities and a vector with element $i$ representing the index of the municipality of respondent $i$ corresponding to row $i$ in the ARD. This returns the posterior samples of $\rho_{km}$ which are subsequently scaled by the size of the groups that were known prior to modelling (known groups) to produce the final estimates. We discuss scaling the posterior estimates by the errors in the estimation for groups of known size in the Supplementary Material. Our code for running the model and scaling is available in our \href{https://github.com/anyarko/subnationalPrev}{subnationalPrev} package.

\section{Simulation studies} \label{sec: sim study}
We run several simulation studies to evidence its ability to produce municipality level estimates and determine how sensitive the results are to the size of the municipalities. 

\subsection{Ability to recover hidden group sizes}
To evidence the ability of our model to produce estimates for multiple municipalities  without incurring an increase in the sample size, we run a simulation study with parameters chosen so that the structure of the simulated data closely resembles the structure of data collected in the Philippines. In total we model six groups with $\mu_{\rho} = (2.5, 3.5, 4.5, 5, 5.5, 6.5)$ so that the first four groups, the known groups, would typically be larger than the unknown groups. 

We then simulate the collection of ARD through a nationwide survey using the synthetic group sizes. We simulate surveys assuming there are $\{5, 10, 15, 20, 30, 50, 100\}$ respondents in the municipalities allowing us to compute the error in our estimates as the sample size varies. Finally, we estimate the size of the hidden groups using both the standard NSUM model and our partially pooled model. Detailed information on the simulation study and plots of the errors can be found in our Supplementary Material.

We first simulate 20 surveys where the number of respondents per municipality is constant and we compute the mean absolute relative error across the municipalities in each survey. For a fixed level of error, our partially pooled model allows us to estimate the simulated number of children being trafficked in each municipality with a smaller sample size than the standard model. The median error in the results of the partially pooled model are lower than for the standard model across all sample sizes with the difference in median error markedly increasing from 15 respondents and more suggesting that the decrease in error for each additional respondent is greater for our model.

For this simulation, the most accurate performance for the standard model was using the maximum 100 respondents per municipality. The partially pooled model achieved a lower level of error with 15 respondents per municipality in all surveys. For the Philippines it took on average 28 minutes to carry out, code and validate each respondent's interview. Reducing the number of respondents in each municipality from 100 to 15 represents a time saving of 5,950 interviewer hours or nearly 250 interviewer days. This does not include the time saved by organising a study with a smaller sample size. 

\subsection{Sensitivity to the municipality size}
Next we allow the number of respondents in the municipalities in a survey to vary. We run 5 surveys of this type and group the results of the municipalities that had the same sample size to determine the range of error for the given sample size when included with other municipalities of varying size. Again the partially pooled model outperforms the standard model at every sample size and has few outlier errors.

We also examine the ability of the partially pooled model to account for varying average counts of the number of people known to each respondent (degree). We simulate two sets of 10 surveys. One where each respondent's degree is 100 and another where this is 800. Both the standard model and partially pooled model are considerably more accurate when the average degree is higher with the median error at lowest number of respondents (5) falling from 1.1 and 0.7 to 0.75 and 0.55 for the standard and partially pooled models, respectively. With a higher average degree the range of errors for the standard model is constant but still large when the number of respondents increases. The partially pooled model's range of errors is much smaller for the larger degree simulations. As it relates to surveying, varying definitions of knowing a member in the groups presented to respondents change the degrees of the network that are being estimated. Our results provide evidence that soliciting information about weak ties thus leading to larger degrees may be more informative in some cases \citep{Granovetter_1973, Feehan_2022}.  Plots of the performance are given in our Supplementary Material.

\section{Trafficking to produce CSEM in the Philippines} \label{sec: results}
We carried out a nationwide survey in the Philippines from October to December 2022.  The sample size in each municipality was 24 individuals, and this was chosen to balance the error in the resulting estimates with the financial cost of the survey. Responses in the municipalities were collected with one barangay in each municipality sampled. A stratified multistage cluster sampling approach was used to determine the number and spatial distribution of barangays to sample for a representative household survey. Municipalities in the Autonomous Region in Muslim Mindanao were excluded due to concerns around terrorism and the safety of survey enumerators.

\begin{figure}
    \centering
    \includegraphics[scale=0.4]{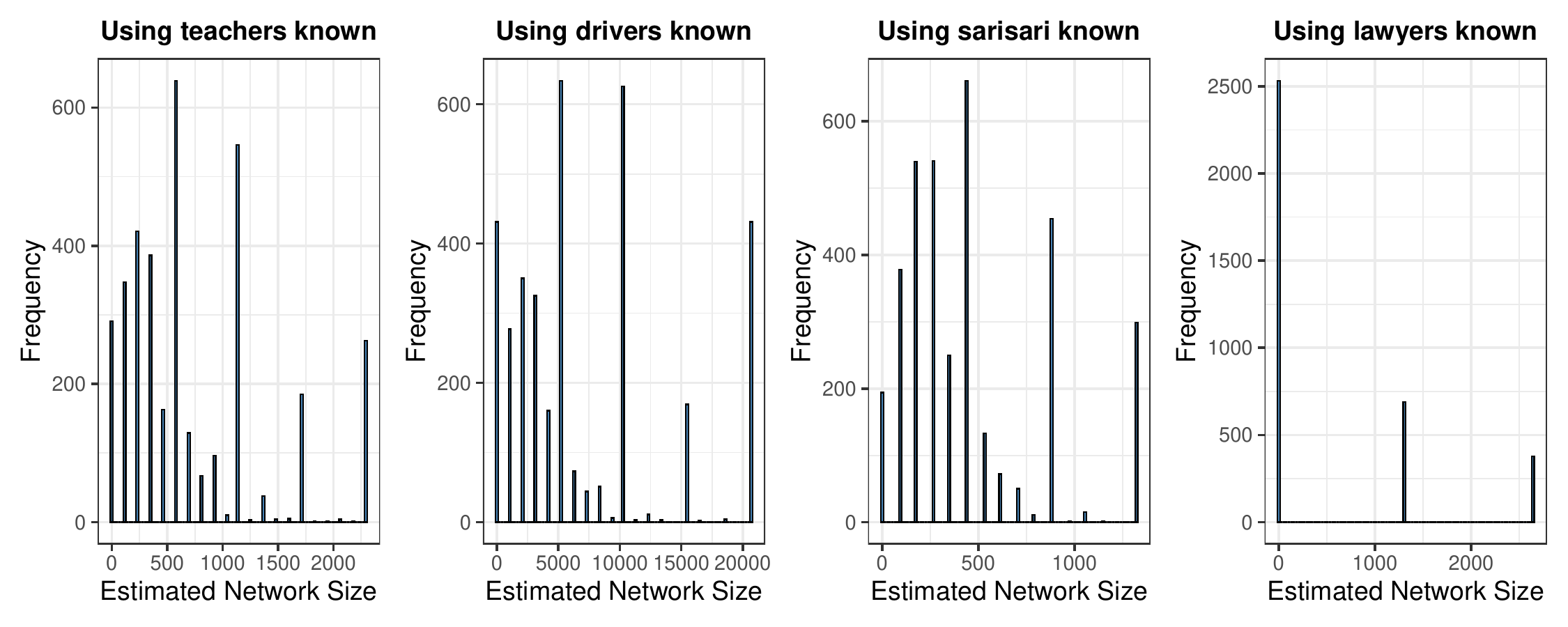}
    \caption{Estimated degree distribution for some known groups. The leftmost three were selected and rightmost is an example of a rejected group.}
    \label{fig:known_group_selection}
\end{figure}

All respondents were asked for basic demographic information about themselves and their household. Given the lack of municipality level demographic data in the Philippines, we estimated the prevalence of the known groups (construction workers, PUV drivers, sari-sari store owners and teachers) by asking the members of their household if they belonged to any of these groups.  The known groups were chosen using a novel procedure of examining the estimated degree distributions visible in Figure \ref{fig:known_group_selection} using the \cite{Killworth_1998b} estimator one group at a time. Known groups were rejected for their lack of ``coarseness'' such as the lawyers known group in Figure \ref{fig:known_group_selection} placing all respondents into only three categories of degree size. The unusual peaks in the distributions are artefacts of respondents at times giving round numbers rather than patiently enumerating their contacts in the queried group and introducing random noise to combat this effect was discussed but not used in this work. We offer more detail and plots in our Supplementary Material. To estimate the size of the hidden populations respondents were then asked about how many people in their municipality they knew who trafficked children to produce CSEM. Survey enumerators received training about CSEM and a strict safeguarding protocol was designed to ensure the safety of both respondents and enumerators. 

Our scoping review in the Philippines and case work from IJM Philippines identified that trafficking to produce CSEM in the Philippines is almost solely familial.  Since our study involves two distinct hidden populations - traffickers (perpetrators) and children being trafficked (victims) - and given that trafficking in the Philippines is predominantly familial, we expected reports of these groups to be correlated. Specifically, if a respondent reported knowing a trafficker, we expected they would also know that trafficker's victims, and vice versa. However, we found that 77\% of the respondents that reported knowing traffickers did not report knowing any victims. Similarly 72\% of the respondents that reported knowing victims, did not report any traffickers. We hypothesised that this lack of reporting is due to our definition of knowing someone in an NSUM survey, as if respondents `knew' the trafficker, may not `know' the victim, in our definition of `knowing'.  Additionally, \cite{Feehan_2022} found that changes in the definition of `knowing' do not necessarily mean an increase in the estimates of the stigmatised population. To mitigate this issue, if a respondent reported knowing one of victims or traffickers, but not the other, we set the value of the missing data to a fixed value. These values were set to the average number of victims per trafficker from casework data (independent of the survey) and the average adult household size respectively. We ran a simulation without the imputation step and the 95\% credible intervals for all but 7 municipalities overlapped. However the mean estimates are higher given that we are adjusting for an under-reporting bias. Nevertheless we find use in the imputation giving the pattern of reporting described.

We fit our model to the 3,600 responses from the national household survey in the Philippines. We ran two chains using the Hamiltonian Monte Carlo algorithm in Stan for 5,000 iterations. The first 3,000 iterations were used to tune sampling parameters by Rstan and subsequently discarded. We examined trace plots to determine an appropriate number of iterations and to ensure that Markov chain had converged. We used the same values for the prior parameters as in the simulation study. Diagnostic plots and information can be found in the Supplementary Material.

\subsection{Estimates for the number of traffickers and children being trafficked}
\begin{figure}[ht]
    \centering
    \includegraphics[width = 0.32\textwidth]{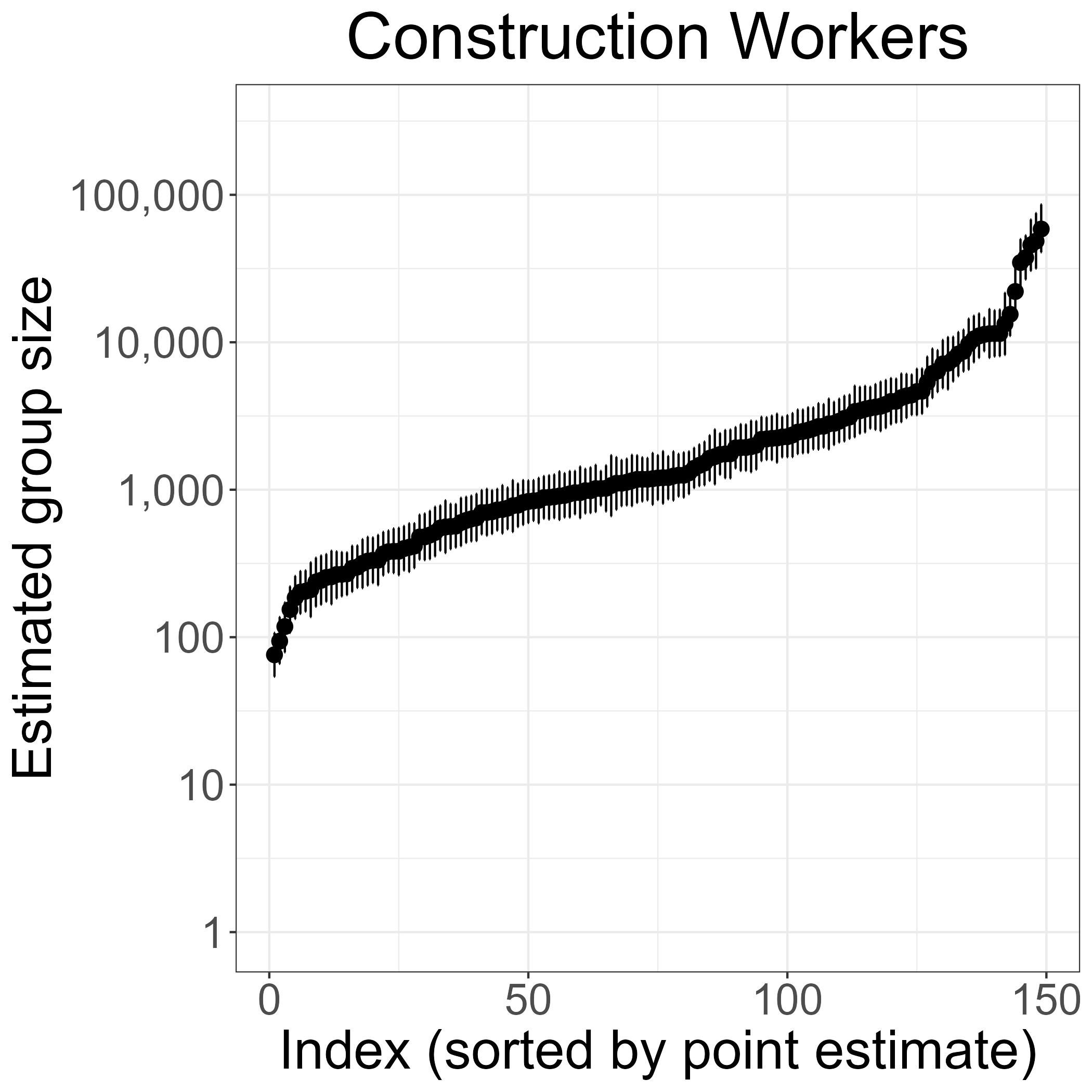}
    \includegraphics[width = 0.32\textwidth]{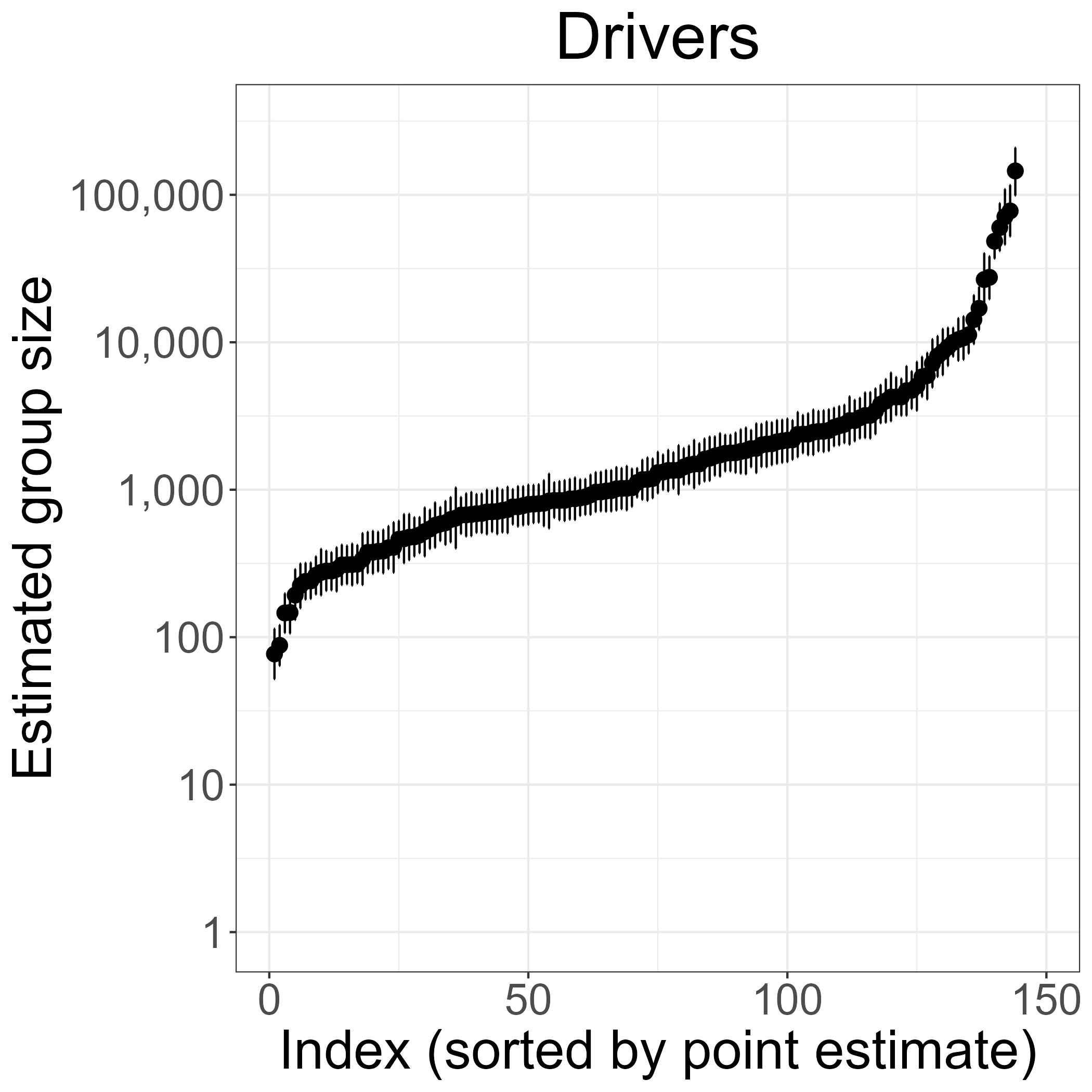}
    \includegraphics[width = 0.32\textwidth]{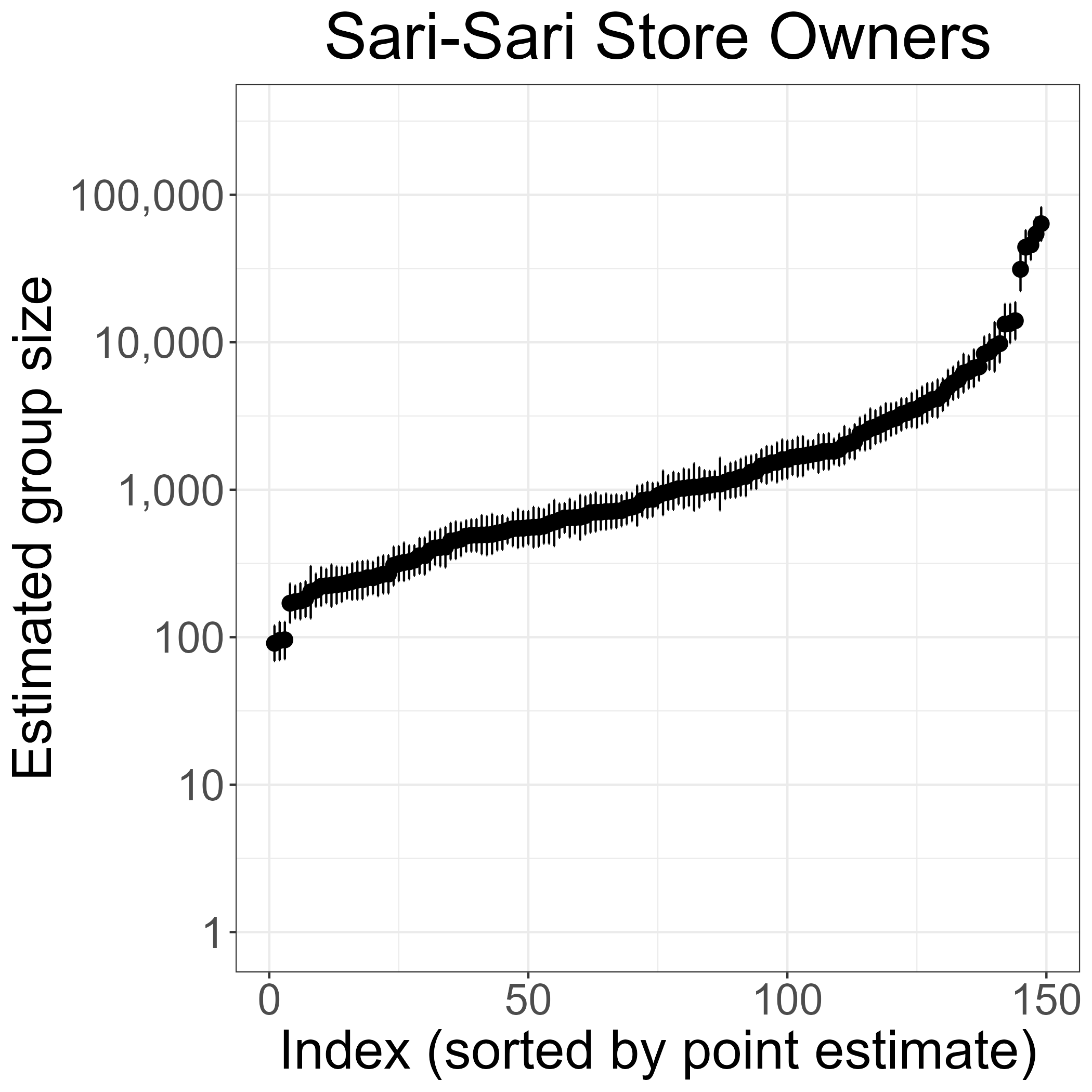}
     \includegraphics[width = 0.32\textwidth]{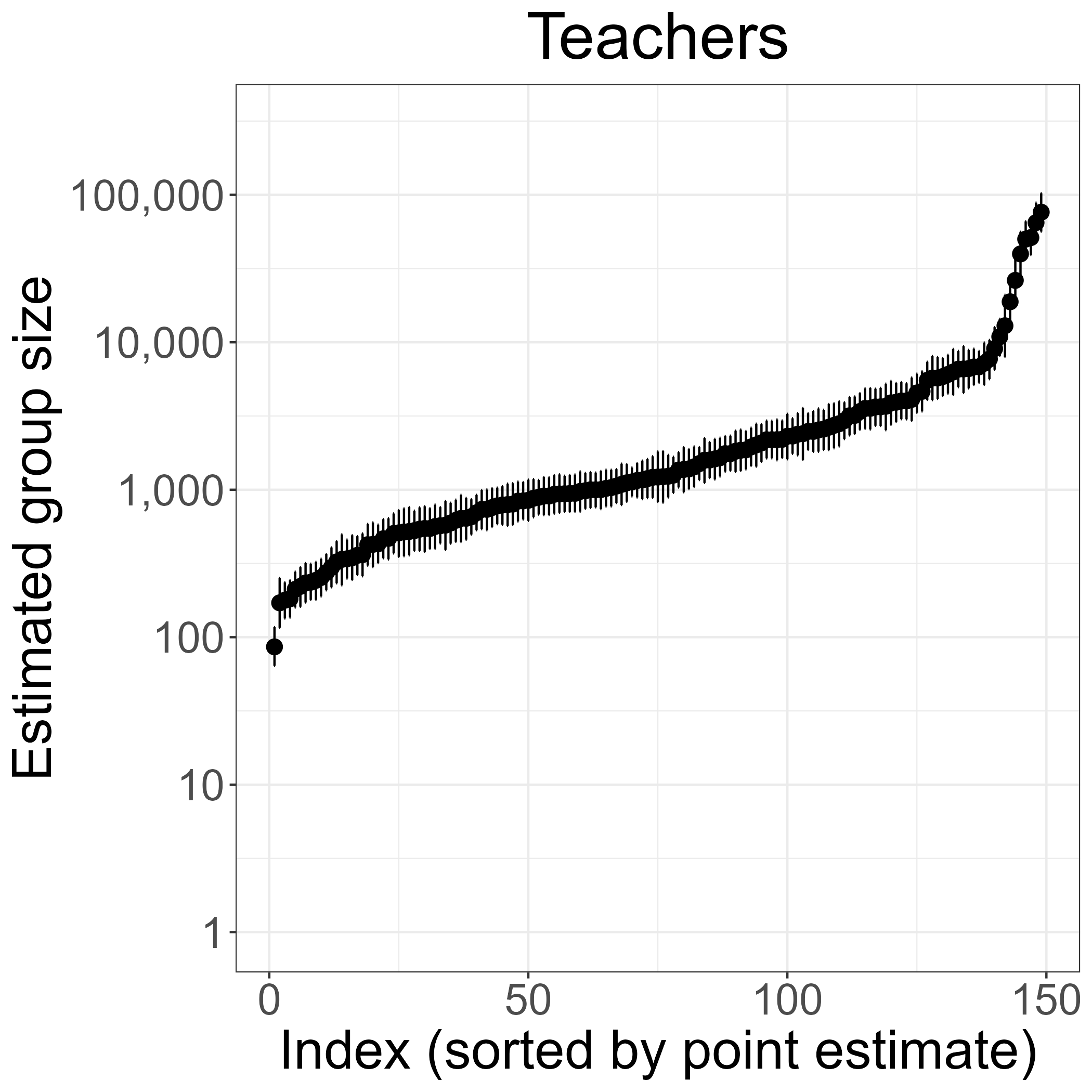}
    \includegraphics[width = 0.32\textwidth]{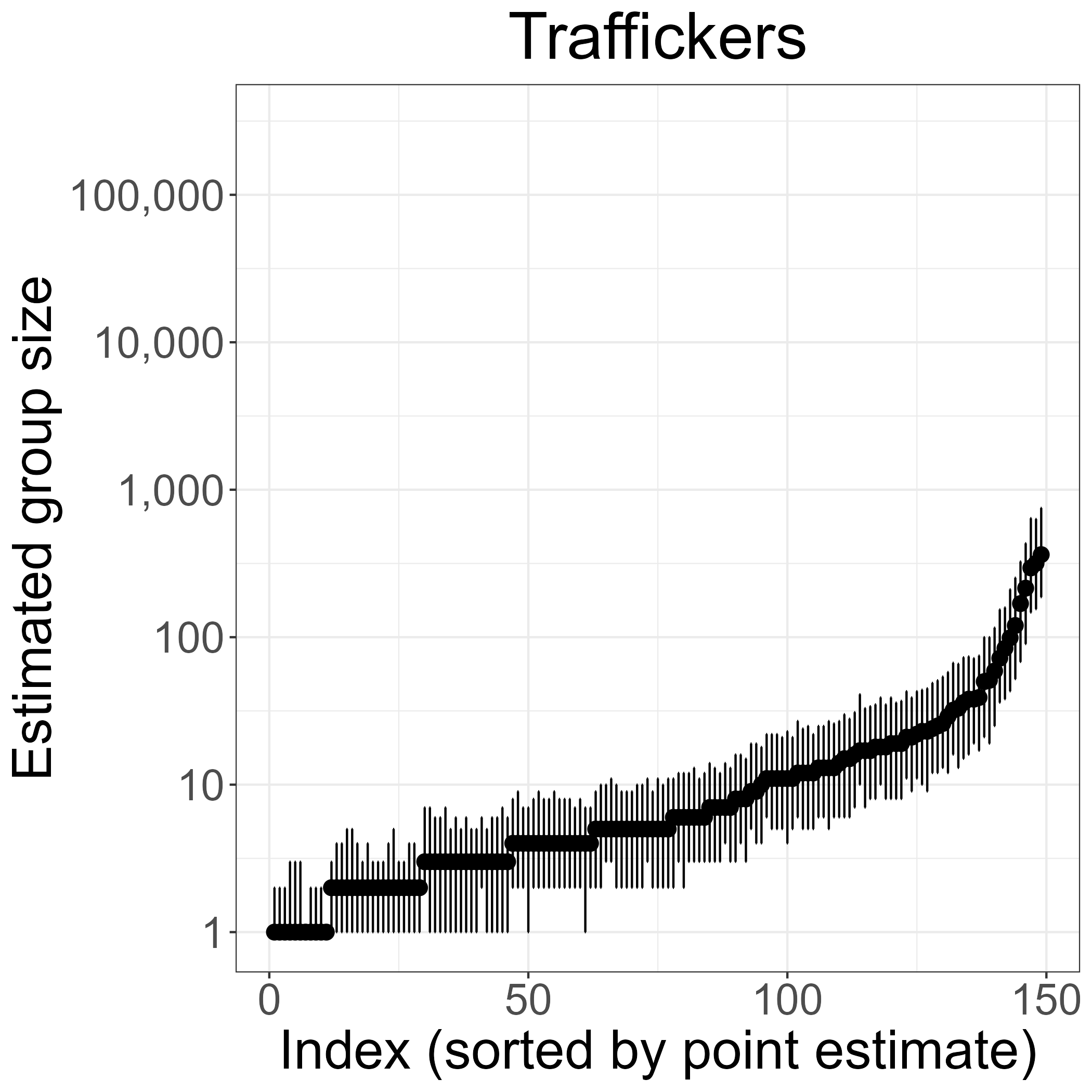}
    \includegraphics[width = 0.32\textwidth]{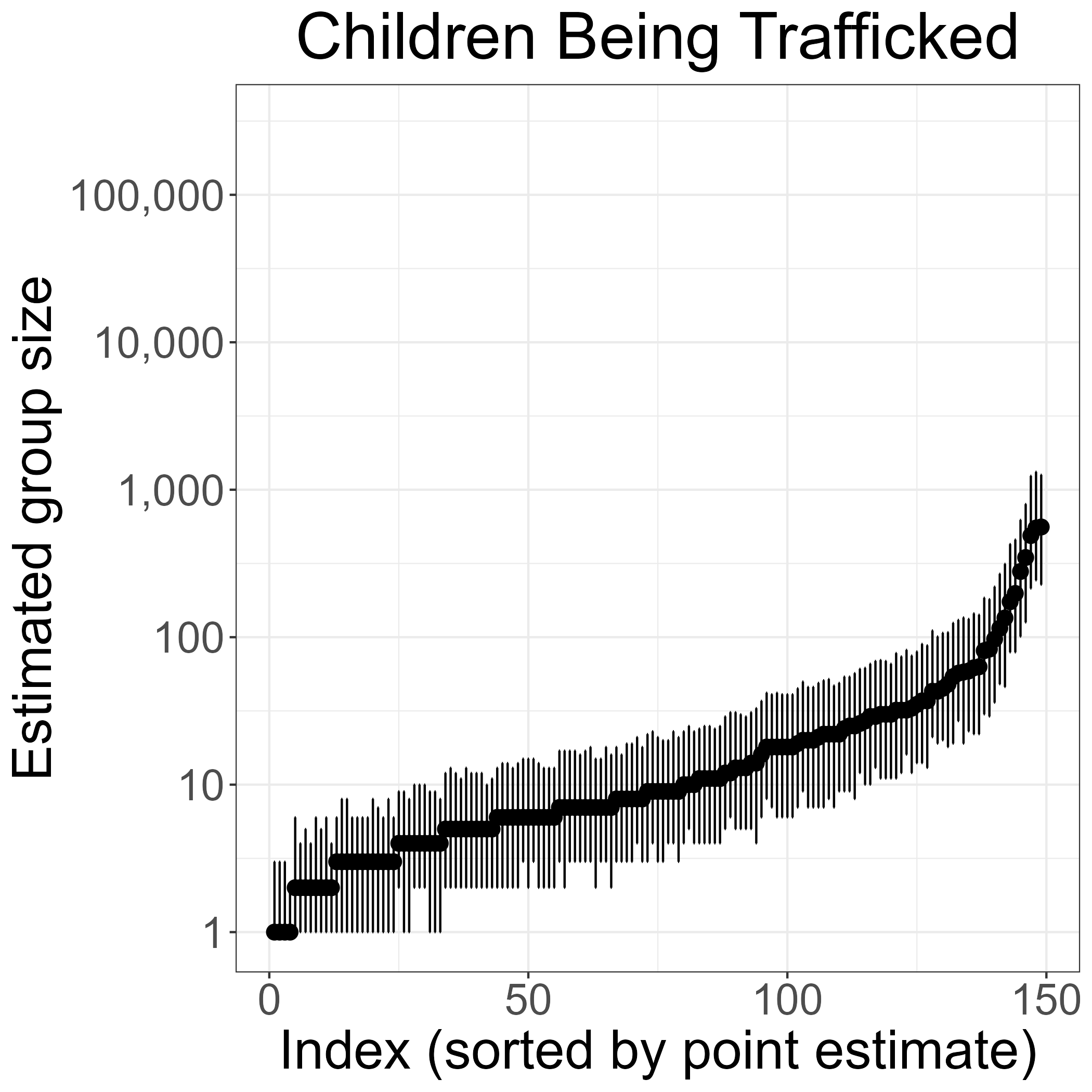}
    \caption{Posterior means for the estimated number of people in each municipality in each (non-zero) group ($N_m\exp(-\rho_{km})$). The grey bars represent 95\% credible intervals. In each plot, the municipalities are ordered from smallest to largest group size. Note the log scale on the $y$-axis.}
    \label{fig:estimated_group_sizes}
\end{figure}

The estimated number of people in each group in each municipality is shown in Figure \ref{fig:estimated_group_sizes}.  For each of the known groups, we find the sizes of the groups vary from the low hundreds to the high thousands. We find there are some municipalities with high numbers of sari-sari store owners and construction workers which correspond to urban and densely populated municipalities. The size of the hidden groups is smaller than the known groups in all municipalities. San Ildefonso and San Vicente municipalities, both in Ilocos Sur province both have the lowest estimate number of children being trafficked, with posterior median 1 child and 95\% credible interval (0, 3). Santa Praxedes has the lowest estimated number of traffickers, with posterior median 0 traffickers [95\% CI (0, 1)]. We estimate that 94 (62\%) of the municipalities have fewer than ten traffickers and 79 (52\%) have ten or fewer children being trafficked. The nine municipalities with the highest number of children being trafficked all have city status. The city of Davao has the highest estimated number of children being trafficked, with posterior median 559 [95\% CI (227, 1262)]. The city of Davao also has the highest estimated number of traffickers with a posterior mean of 363 [95\% CI (187, 752)]. 

\begin{figure}
    \centering
    \includegraphics[scale=0.85]{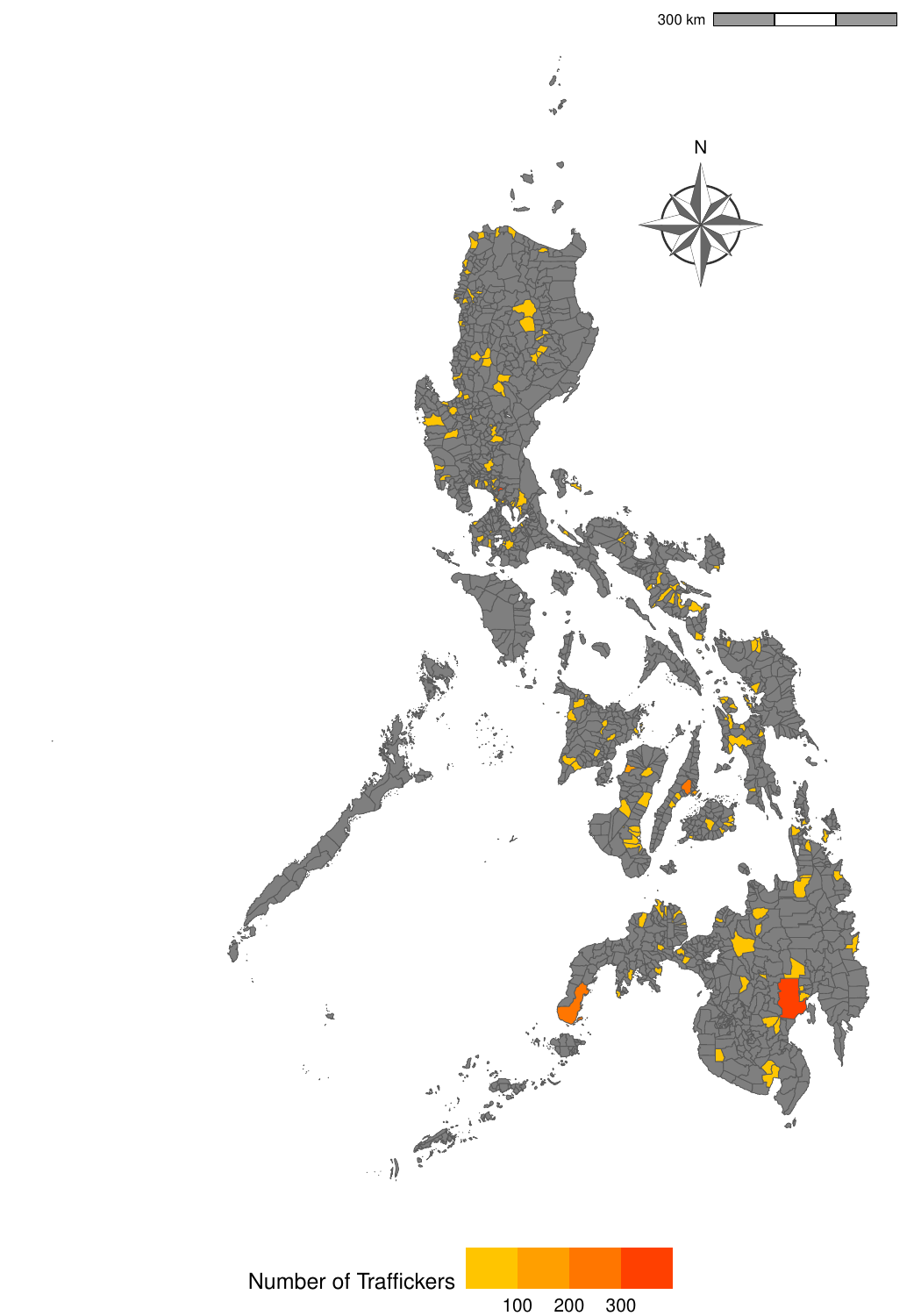}
    \caption{The estimated number of adults trafficking children to produce child sexual exploitation material across the sampled municipalities.}
    \label{fig:traffickers}
\end{figure}

Figure \ref{fig:traffickers} shows a map of the Philippines where the municipalities in the study are shaded by the estimated number of adults trafficking children to generate online CSEM (the corresponding map showing the estimated number of trafficked children in each municipality is in our Supplementary Material). Municipalities coloured in red have the highest number of traffickers. This map can be used by social service providers, law enforcement agencies, and third sector organisations to identify municipalities where interventions should be prioritised.  From this, we can see the municipalities with the highest number of adults trafficking children are concentrated in urban areas, particularly around Manilla, Davao and Cebu City. In the more rural parts of Luzon and Mindanao islands, the absolute number of traffickers in each municipality is lower than in the urban parts. Across the sampled municipalities, we estimate there are a total of 5,042 [95\% CI (1998, 11810)] children being trafficked and 3,077  [95\% CI: (1412, 6226)] traffickers.

\begin{figure}
    \centering
    \includegraphics[scale=0.4]{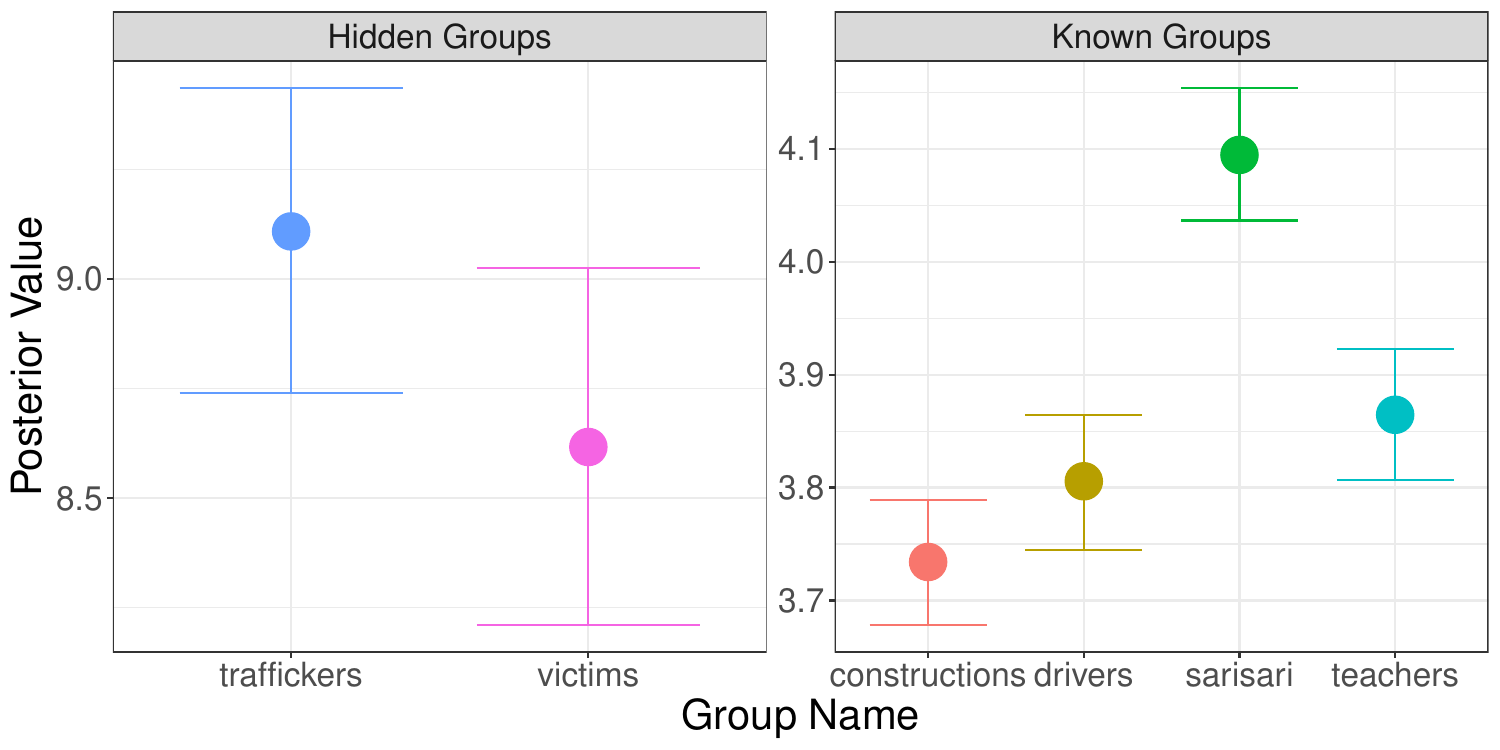}
    \caption{Posterior means for $\mu_{\rho_k}$ the mean of the population occurrence across municipalities. 
    The bars represent 95\% credible intervals.}
    \label{fig:mu_rho_posterior_means}
\end{figure}

Figure \ref{fig:estimated_group_sizes}  shows the uncertainty in the prevalence estimates. The size of the uncertainty in the estimates are small enough for us to make meaningful conclusions about the hidden groups in each municipality. For the known groups, the standard deviation of the posterior mean was 0.3. This suggests that inter-municipality variation of population occurrences is low. Our partially pooled model uses parameters that represent the mean prevalence across municipalities. One interpretation of these parameters corresponds to the national distribution of groups sizes across municipalities, this is shown in Figure \ref{fig:mu_rho_posterior_means}. 

\begin{table}
\caption{Posterior mean estimates of model parameters across all municipalities and respondents.\label{tab:implied_population_parameters}}
\begin{tabular}{|c|c|c|c|c|c|c|}
\hline
Parameter & PUV drivers & Construction Workers & Sari-Sari & Teachers & Traffickers & Children\\
\hline
$\mu_k$ & 5.4438 & 5.8472 & 4.0767 & 5.1323 & 0.0271 & 0.0443\\
\hline
$w_k$ & 3.3985 & 1.5861 & 33.8608 & 3.8706 & 0.0195 & 0.0105\\
\hline
\end{tabular}
\end{table}

Table \ref{tab:implied_population_parameters} shows the expected number of people known in each group, denoted $\mu_k$, and respective overdispersion across all respondents. All groups, both known and hidden, show levels of overdispersion relative to a Poisson distribution, with Sari-Sari store owners showing the lowest level of relative overdispersion. The considerable level of overdispersion in the other groups justifies our use of the Negative Binomial. 

As no current estimates for the prevalence of this crime exist in the Philippines -- either at national or municipality level -- validating our results is difficult. However, we can compare our findings to related work. The Global Partnership to End Violence against Children and UNICEF's \textit{Disrupting Harm} survey in the Philippines \citep{unicef2022} was a nationwide survey in the Philippines, to estimate how many children had experienced online child sexual exploitation and abuse. Their definition of abuse was much broader than ours, including blackmail or receiving illicit images. The \textit{Disrupting Harm} estimated prevalence directly for the respondents and did not collect ARD. The survey found that around 2 million Filipino children had suffered online child sexual exploitation and abuse on average, around 1,400 children per municipality. Our estimates were around a tenth of the size of those from \textit{Disrupting Harm}, and given that we were estimating one of the most extreme forms of abuse \textit{Disrupting Harm} considered, it is plausible that there is agreement between our two estimates. 

\section{Discussion} \label{sec: discussion}
We have developed a partially pooled variant to the NSUM models and demonstrated it on data we collected in the Philippines on trafficking to produce child sexual exploitation material. Our model allows us to estimate the size of hidden groups in multiple geographical areas without increasing the sample size of a survey. 

We analysed a novel data set on trafficking to produce CSEM in the Philippines. This is the first time estimates for trafficking to produce CSEM have been produced in the Philippines at municipality level. Across the municipalities surveyed, we estimate there to be around 3,000 traffickers and 5,000 children being trafficked. We found the City of Davao has the highest number of children being trafficked as well as traffickers. In around half of the municipalities we surveyed, there were at least ten children being trafficked.

We found that parameterising the Negative Binomial distribution through its mean and overdispersion was a flexible distribution for modelling both known and unknown populations. This is due to its direct modelling of overdispersion which is suited to groups that are not equally spread in the population. In our results, we found considerable overdispersion in the majority of groups, suggesting that the Poisson distribution is not suitable even for modelling known groups.  While the methods proposed by \citet{Laga_2023} offer valuable insights into bias estimation within social networks, our partially pooled NSUM model builds upon these foundations to provide a more scalable and flexible solution for estimating hidden populations across multiple geographical areas. By carefully choosing our distributions and modelling strategies, we address the limitations of previous approaches and enhance the robustness of our estimates, making our model well-suited for complex, multi-regional studies like our investigation into child trafficking in the Philippines.

Our results, and the research conducted in the development of our model illuminated various areas for further research, advocacy, and programming to further address trafficking to produce child sexual exploitation material. These include (1) the promotion of community-based reporting to prevent trafficking for the production of CSEM; (2) the deployment of strong community education and awareness initiatives; (3) intensifying  law enforcement efforts to apprehend and prosecute traffickers and demand-side offenders; (4) improving financial transaction monitoring  to detect and report suspicious transactions related to child sexual exploitation, blocking those linked to offenders and traffickers; (5) promoting the involvement of those with lived experience into research on CSEM to better understand the behaviour of perpetrators and traffickers. In September 2024, the President of the Philippines, Ferdinand R. Marcos Jr., cited our research during an CSEM summit \citep{Marcos24}. This follows a new presidential directive he signed into law in August 2024, setting up the Presidential Office for Child Protection that  will tackle crimes including online child sexual abuse. 

Our model has the potential to be applied to a wide range of sociological issues. Although we collected aggregated relational data on trafficking of children, our model can, in theory, be applied to estimate the size of any hidden group, such as those at risk of modern slavery, those who use illegal drugs or people with stigmatised sexually transmitted diseases. Work with relevant stakeholders and practitioners who work with the hidden group should be consulted to ensure that the model is responsibly applied. Our model is also not restricted to national studies. It can be applied to the geographical unit most suited for the hidden group or planned intervention. For example, our model can be applied to a city, where estimates are required for the size of the hidden population within different neighbourhoods. 

The imputation of counts for people who partly reported CSEM is a novel process in ARD related work. We see future application to pairs of groups that face a similar lack of reporting but are linked in definition such as intravenous drug users and their providers however more future work on how this should be approached would improve the understanding of what can reliably be said about the imputed networks. Based on there being more non-zero counts of the hidden population after imputation, one would expect the estimates given to be overestimates. For this study of hidden populations however, we could not account for respondents who knew of instances of trafficking to produce CSEM and reported neither traffickers or victims; a form of transmission bias whose effect is potentially larger. Previous ARD work has introduced the idea that the estimates of groups that are correlated should be scaled in the same way \citep{Laga_2023}. This could be extended to reports of groups that are linked by definition. 

These hidden population counts may fall into a larger class of data missing not at random. Instrument variables have been used in a related way to account for selection biases. In a nonparametric causal inference based framework given by \cite{Tchetgen_2017} three requirements are given for an instrument variable which include predicting the nonresponse process and independence from the outcome being modelled. Instrument variables maybe a useful addition to improve the robustness of our estimation but confirming these requirements may need a more detailed examination of the surveyed personal networks which is complicated by the criminal element of our hidden groups hence a simpler approach guided by case data was chosen.

We chose not to model transmission bias as in typical NSUM studies due to the difficulty of estimating it from just ARD data. Transmission is typically modelled as a multiplier, $\tau_k$, to the prevalence of each hidden group. In our model this would result in an identifiability problem since the value of the likelihood would not be distinguishable between groups that are for example 10\% of the population with 50\% being visible in networks and reported and groups that are 50\% of the population but with 10\% reported. 

The Game of Contacts \citep{Salganik_2011} was proposed as a way to estimate transmission thus placing it outside the scope of parametric modelling however is requires some of the respondents in the survey to be part of the hidden groups which is difficult for studies of groups where crime is a factor. For \cite{Maltiel_2015}, when trying to model transmission the lack of an informative prior lead to large uncertainty in the estimates of prevalence. As we did not have contact with the hidden groups outside of discussions with survivor consultants, the effect of this bias may be best represented by a multiplier to the results of our study as recommended in a review of using NSUM specifically to estimate trafficking by \citet{Shelton_2015}.

One limitation of our model is that it does not include respondent bias, which has been included in other NSUM models \citep[see, for example,][]{Teo_2019}, or \citet{Maltiel_2015}. We only include the degree of each respondent. Introducing a respondent bias term would allow for heterogeneity within social networks within each municipality. However, given that we are already modelling a heterogeneous structure within the Philippines, further work would be required to introduce municipality level heterogeneity. As the size of the groups in each municipality differ, the bias would need to scale according to these sizes, which current methods do not currently allow.

Further work could consider spatial correlation models, for example by placing a joint prior distribution on the municipality level group size parameters. This may further reduce the sample size required for accurate estimates. Additionally, it may be fruitful to generalise the model to produce national and municipality level estimates simultaneously. There is also considerable potential in analysing the data from the Philippines further; this includes in understanding and quantifying the drivers of trafficking and producing counterfactual analysis to further determine interventions. 

One way to produce a national estimate from the municipality level estimates would be to generate out--sample predictions for municipalities that were not sampled in the national survey. Further work to determine suitable covariates and a modelling structure for out of sample prediction is ongoing. A method to construct a national estimate by aggregating municipality level estimates could be developed using sampling weights and out of sample prediction. Another method would be to use sampling weights to aggregate the municipality estimates into a national estimate. Several existing methods could be extended to achieve this. In \cite{Gelman2020}, the authors  propose a Bayesian hierarchical model for including sampling weights in surveys, including small area estimation. We could use this post-stratification method with our ARD to municipalities in the Philippines, computing their contribution to the overall national estimate. Another method has been proposed in \cite{Gunawan2020}, which combines the survey weights in the MCMC algorithm. We could use survey weights to generate a representative sample of each municipality and embedding these samples in the MCMC algorithm. This would allow us to sample from the posterior distribution for the estimate in each municipality.

\section{Acknowledgements}
This work was funded by the International Justice Mission's Centre to End the Online Sexual Exploitation of Children and supported by a University of Nottingham Research Fellowship (2021-2023) and a UKRI Future Leaders Fellowship [MR/X034992/1].

\bibliographystyle{rss}
\bibliography{nsum.bib}

\end{document}


\maketitle

\def\spacingset#1{\renewcommand{\baselinestretch}%
{#1}\small\normalsize} \spacingset{1}

\section{Deriving the Negative Binomial model}
In the partially pooled NSUM model, we assume that there are $M$ municipalities and each respondent $i$ reports the number of people they know in group $k$ in municipality $m$, denoted $y_{ikm}$. Our aim is to estimate the number of people in the hidden group $k$ in municipality $m$, $N_{km}$, where the total population size of $m$ is $N_m$. To model this, we follow a method proposed by \citet{Zheng_2006} that allows for overdispersion to be directly modelled. The authors start by proposing a Poisson model with a rate at which the hidden group is known, and allow this rate the vary between respondent and group. Integrating over this rate parameter, brings about a Negative Binomial model. 

We assume that the rate at which respondent $i$ is aware of people in group $k$ in municipality $m$ is given by $\lambda_{ikm}$. Focusing modelling on this rate is a more flexible approach than starting with the assumption that $y_{ikm}$ is a Binomial variable therefore each person in respondent $i$'s network has the same probability of being in group $k$.  We model the number of people in group $k$ in municipality $m$ that person $i$ knows given this rate by
$$
y_{ikm} \mid \lambda_{ikm} \sim \hbox{Po}(\lambda_{ikm}). 
$$
Following \cite{Zheng_2006}, we assume that $\lambda_{ikm} \sim \Gamma(w_k, \beta_{ikm})$, where $w_k$ and $\beta_{ikm}$ are the shape and scale parameters of the distribution respectively. Therefore the probability that  respondent $i$ knows $j$  people in group $k$ in municipality $m$ is given by marginalising over $\lambda_{ikm}$ 
\begin{align*}
    \pi(y_{ikm} = j ) &=\int_0^\infty \pi(y_{ikm} = j \mid \lambda_{ikm}) \pi(\lambda_{ikm}) d\lambda_{ikm}\\
    &=  \begin{pmatrix}
    j + w_k - 1 \\ j
\end{pmatrix} 
\left(\frac{1}{1+\beta_{ikm}} \right)^j \left(\frac{\beta_{ikm}}{1+\beta_{ikm}} \right)^{w_k},
\end{align*}
which is the mass function of the Negative Binomial distribution. This has expectation $\mathbb{E}(y_{ikm}) = \frac{w_k}{\beta_{ikm}}$. Denoting this expectation as $\mu_{ikm}$ allows us to define the variance as $\hbox{Var}(y_{ikm}) = \mu_{ikm} + \frac{\mu_{ikm}^2}{w_k}$. The parameter $w_{k} > 0$ controls for the dispersion in knowing people in group $k$. We then write this distribution as
 \begin{equation}
    y_{ikm} \sim \hbox{NegBin}\left(\mu_{ikm}, w_{k}\right),
\end{equation}
and use a link function to connect the mean and in future work this could include respondent and municipality covariates that allow for out-sample estimations of prevalence. The challenge of out-of-sample predictions is that for each of the municipalities we sampled in the Philippines we were also able to collect distinct data to estimate the prevalence of the known groups which could be used for scaling.  Out of sample municipalities would not have this data available however an analysis of the scaling effect of the municipalities that were sampled could be helpful in scaling the estimates of such municipalities.

\section{Selection of Known Groups} \label{sec: choosing kgs}
We would like to introduce a novel method for selecting the known groups to use in the estimation of groups sizes. ARD studies will typically ask about a number of candidate known groups that are similar to the hidden population and will therefore improving the scaling of the unknown group estimates. The use of these known groups is typically in estimating the degree of each respondent so by extension good known groups are ones that adequately estimate respondent degrees. Previous studies have used the agreement of a leave-out-one approach with the known groups to identify which known groups are good \citep{Habecker_2015}. Although this has the great benefit of being a self-consistent approach, we don't necessarily see this as the approach that gives more accurate estimates; only known groups whose estimates agree which each other when alternatingly excluded.

\begin{figure}
    \centering
    \includegraphics[width=16cm, height=20cm]{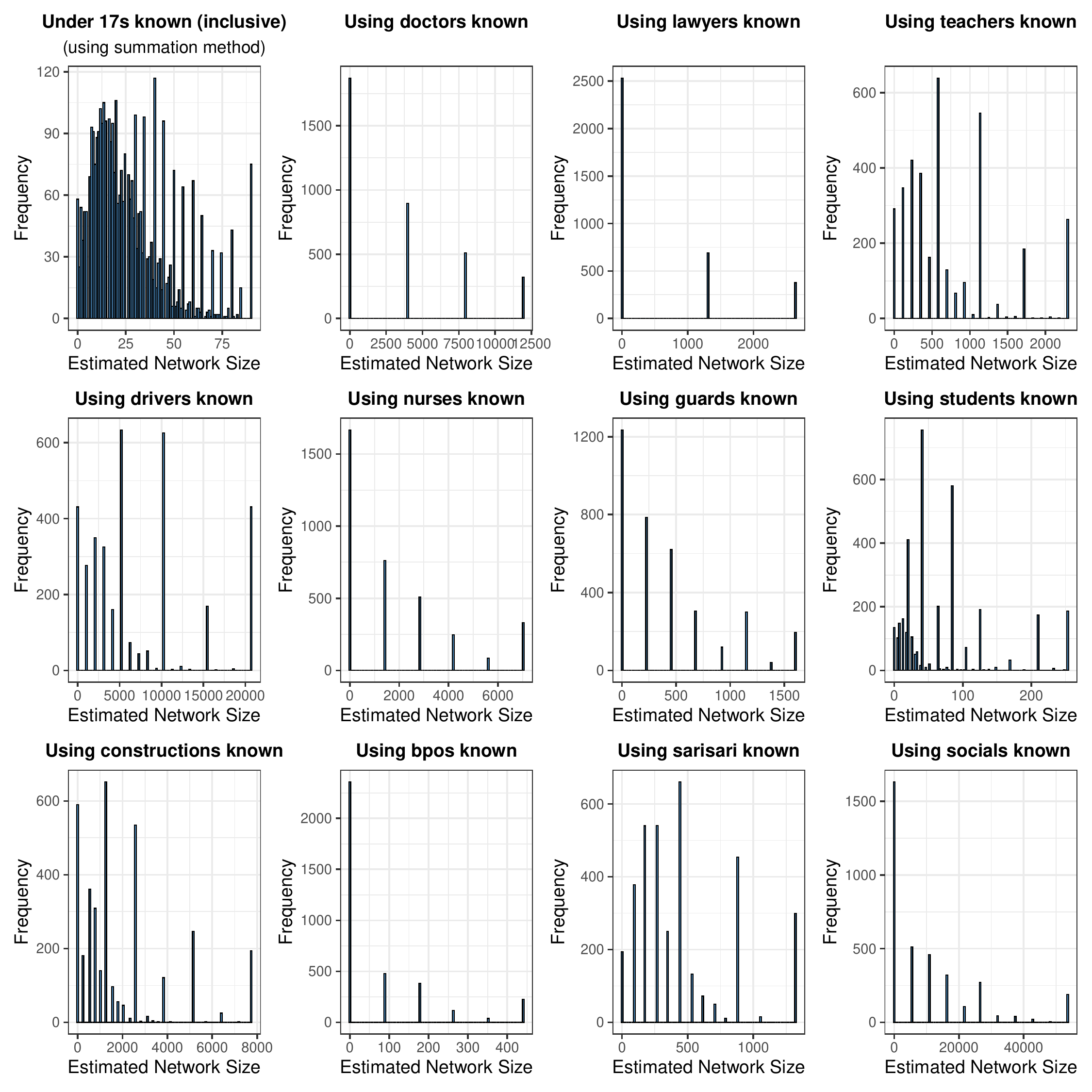}
    \caption{The estimated degree for each respondent using a basic degree estimator}
    \label{fig:degree_estimation}
\end{figure}

Our novel method involves direct observation of the degree distributions the individual known groups estimate and then selecting groups that perform well in this sense. We use one estimator of degree for a single known group given by \citet{Killworth_1998b}. This is $\hat{d_i} = N_m*\frac{y_{ik}}{N_{km}}$. A plot showing the results of using this estimator on each of our known groups is given in Figure \ref{fig:degree_estimation}. In the main paper we mention that groups are dropped for their lack of ``coarseness'' such as the lawyers group. The addition of each new group is computationally expensive. Although any group contain information it is advantageous to use a procedure that prioritises the most informative groups.

\FloatBarrier 
\newpage

\section{Simulation study} \label{sec: sim study}
To evidence the ability of our model to produce estimates for multiple municipalities  without incurring an increase in the sample size, we run two simulations with parameters chosen so that the structure of the simulated data closely resembles the structure of data collected in the Philippines.  We simulate the number of children trafficked to produce CSEM and the number of traffickers in each of the municipalities as well as the size of four known groups. The known groups namely: 
\begin{inparaenum}[1)]
\item construction workers,
\item Public Utility Vehicle (PUV) drivers,
\item sari-sari store owners, and
\item teachers.
\end{inparaenum} 
These correspond to the selected known populations that were collected in the Philippines. We use the plausibility of the estimated degree distributions for each group to choose these known groups from a larger pool of groups. Details of this approach are found in Section \ref{sec: choosing kgs}. In total we model six groups with $\mu_{\rho} = (2.5, 3.5, 4.5, 5, 5.5, 6.5)$ so that the first four groups, the known groups, would be typically be larger than the unknown groups. These values are chosen to ensure the simulated data is similar in structure to the data we collected in the Philippines and also demonstrates our model's ability to capture populations that have a mean occurrence of between 8\% and 0.1\% of the population. Our choice of $\sigma_{\rho_{k}} = 1$ for all groups extends the range of group sizes that can be modelled into 0.01\% of the population.  For the remaining model parameters, we set $w = (35, 35, 35, 35, 2, 0.3)$ to approximate situations where the variance of the known groups is close to the mean and when the variance is higher than the mean as we expect when modelling ARD of stigmatised populations. We set $\mu_\delta = 5.5$  and $\sigma_\delta = 1$ similar to previous ARD simulation studies \cite{Maltiel_2015, Laga_2023}. 

When fitting the model we set the prior standard deviation parameters $\sigma_{\mu_{\rho}} = 10$ and $\tau = 1$ to reflect our belief that the mean size of groups may vary significantly but given a group the inter-municipality variation will be well modelled by a half standard normal. For the overdispersion parameter $w_k$, we set the standard deviation of the half-normal prior distribution to be $\sigma_w = 10$ so that dispersion can be modelled for groups that have high and low values for the first parameter of our negative binomial model. These prior distributions have positive support and allow us to make either vague or specific assumptions to be made about the national prevalence of the hidden group. For the log of the number of people in respondent $i$'s network $\delta_i$, we choose the mean of the distribution used by \citet{Maltiel_2015} and fix $\mu_\delta = 5.5$ due to an identifiability issue. We set $\sigma_\delta \sim \mathcal[0, 1.5]$. The uniform prior distribution on $\sigma^2_\delta$ restrict the variance parameter from taking unreasonably large parameter values \citep{Gelman2006}. 

We then simulate the collection of ARD through a nationwide survey using the synthetic group sizes. We simulate surveys assuming there are $\{5, 10, 15, 20, 30, 50, 100\}$ respondents in each municipality. This is to allow us to compute the error in our estimates as we decrease the sample size. Finally, we estimate the size of the hidden groups using both the standard NSUM model and our pooled model. 

\subsection{Average error across municipalities}
\begin{figure}[ht]
    \centering
    \includegraphics[width=0.55\textwidth]{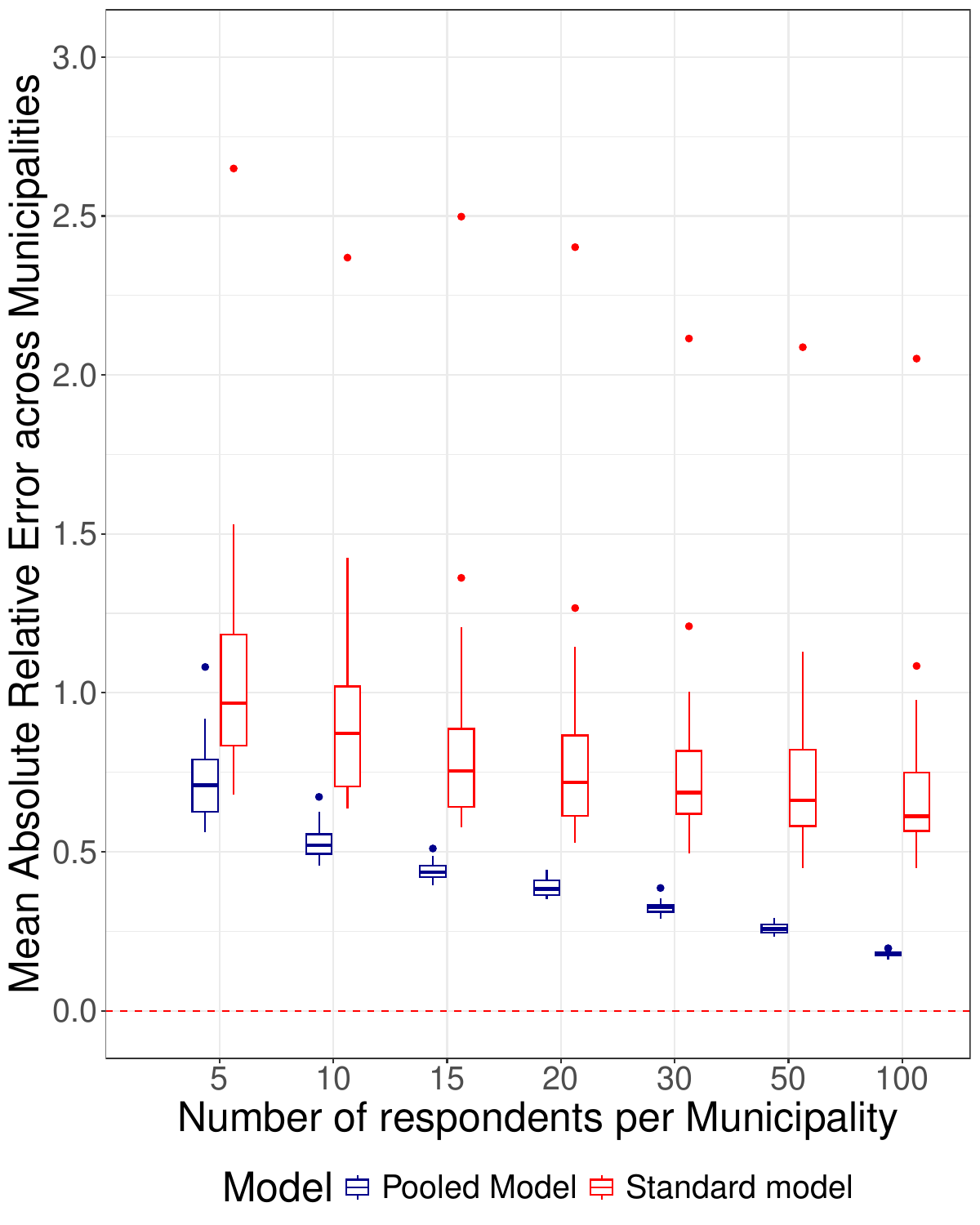}
    \caption{Box plots showing the mean absolute relative errors across groups across the synthetic municipalities simulated against the sample size of every municipality in each run.}
    \label{fig:boxplot_region_errors}
\end{figure}

 Firstly for each sample size we simulate 20 surveys. For each simulated survey, we record the mean error in each municipality then take the average across the municipalities shown Figure \ref{fig:boxplot_region_errors}. Our pooled model outperforms the standard model for all sample sizes. For a fixed number of respondents, the error using our pooled model is always lower than the error using the standard model, except when using 5 or 10 respondents, when there is some overlap likely due to random error. It is also noticeable that the range of errors for our model is smaller than the standard model. For example, when using 100 respondents per region, in all simulations the error for our pooled model was around 0.15, whereas for the standard model it ranged from 0.47 to 2.1. 

 For a fixed level of error, we can see that our partially-pooled model allows us to estimate the simulated number of children being trafficked in each municipality with a smaller sample size than the standard model.

 \subsection{Average error varying sample size}
\begin{figure}
     \centering
     \includegraphics[width=0.6\linewidth]{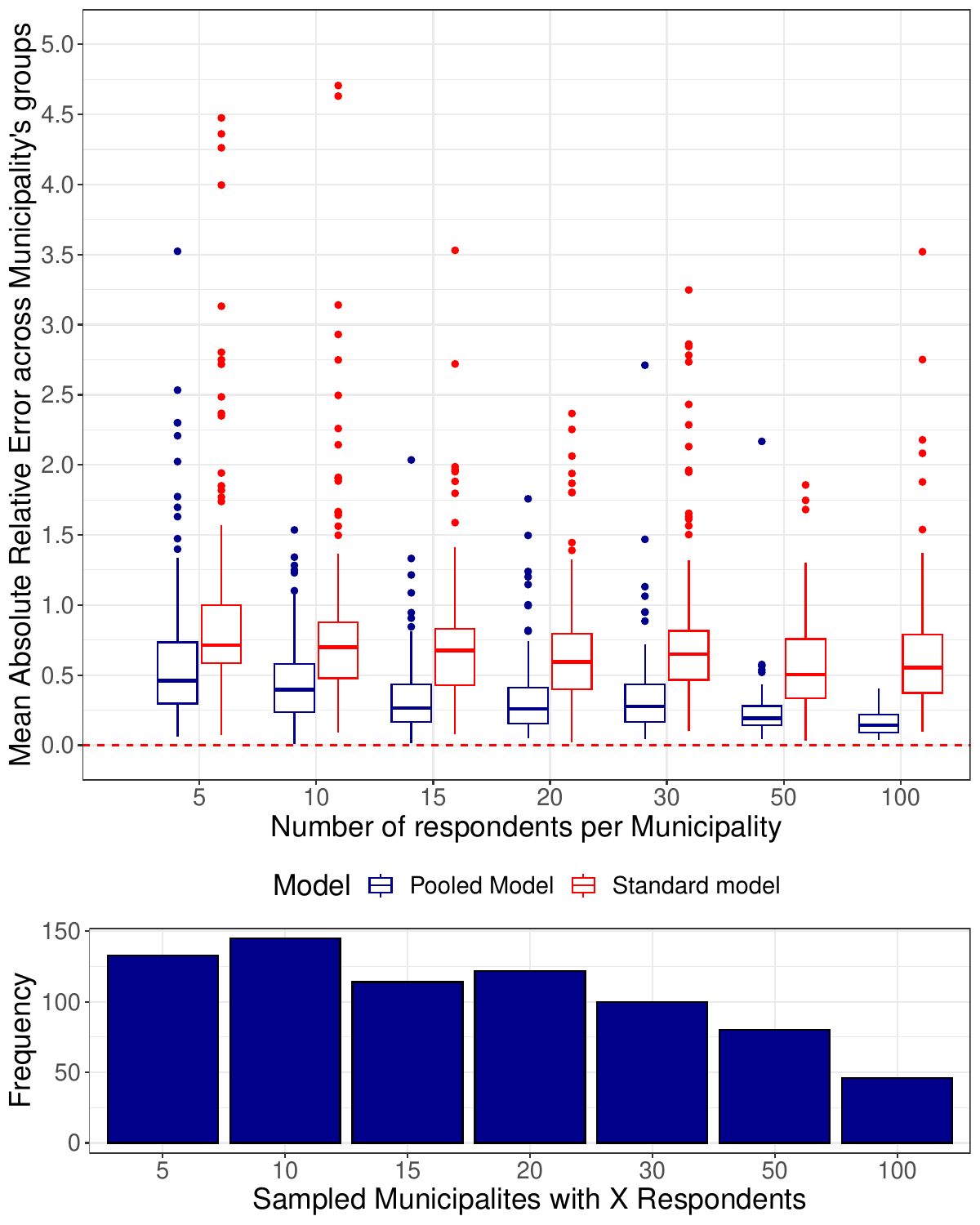}
     \caption{Top: Boxplots showing the mean absolute relative error across groups in each municipality for municipalities with different numbers of respondents and across 5 simulations. Bottom: The number of times a municipality of each sample size was drawn across the 5 simulations.}
     \label{fig:uneven_respondent_numbers}
 \end{figure}

 Our study was able to obtain the required sample size for each municipality. This may not be the case for future studies and result in uneven respondent numbers across the sampled municipalities. To test the model in this scenario we simulated data for 150 municipalities with the number of respondents per municipality randomly selected from {5, 10, 15, 20, 30, 50, 100} and with the condition that the sum of the respondents equals 3600 (as in our Philippines data). We opted for the same set of possible respondent numbers as the previous simulation for comparability. What we wanted to examine was the difference in uncertainty across municipalities with different sample size therefore we calculated the average error in the estimation of groups for each municipality and plotted this for 5 sets of 150 respondent sample numbers with each one representing a survey. We show this in Figure \ref{fig:uneven_respondent_numbers} and we also provide the number of times a municipality with each sample size  occurred in the five different simulations.

Figure \ref{fig:uneven_respondent_numbers} shows that the pooled model again outperforms the standard NSUM model with fewer extreme errors for all municipality respondent sizes. The pooled model also has a distinctly smaller interquartile range of errors at higher numbers of respondents. The pooled model's performance around 24 municipalities is again better than the standard model and justifies its use on the Philippines data.

\subsection{Small and large degree size}
\begin{figure}
    \centering
    \includegraphics[width=0.4\linewidth]{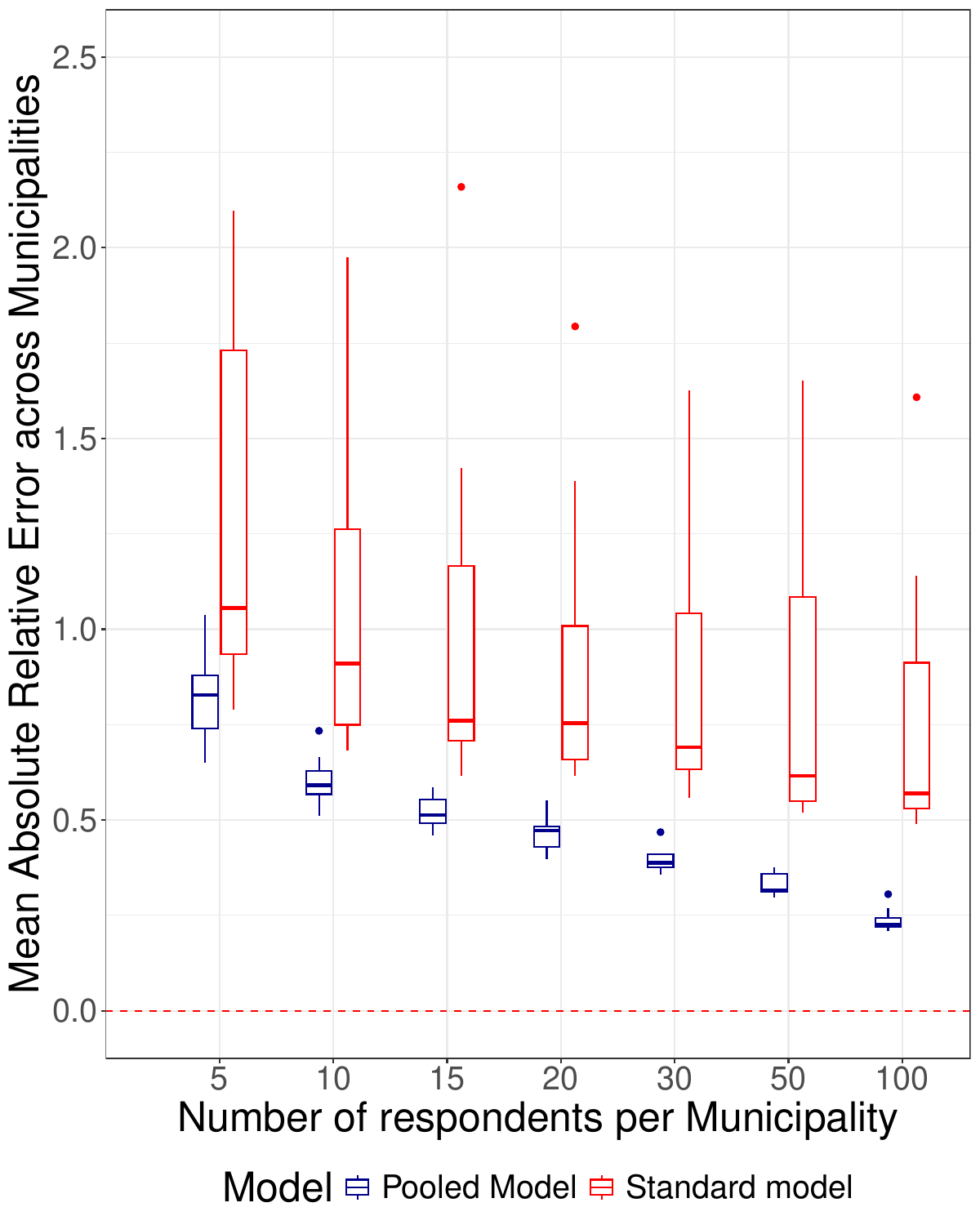}
    \caption{Boxplots showing the MARE. All respondents have a degree of 100 in this data. }
    \label{fig:degree_100}
\end{figure}

\begin{figure}
    \centering
    \includegraphics[width=0.4\linewidth]{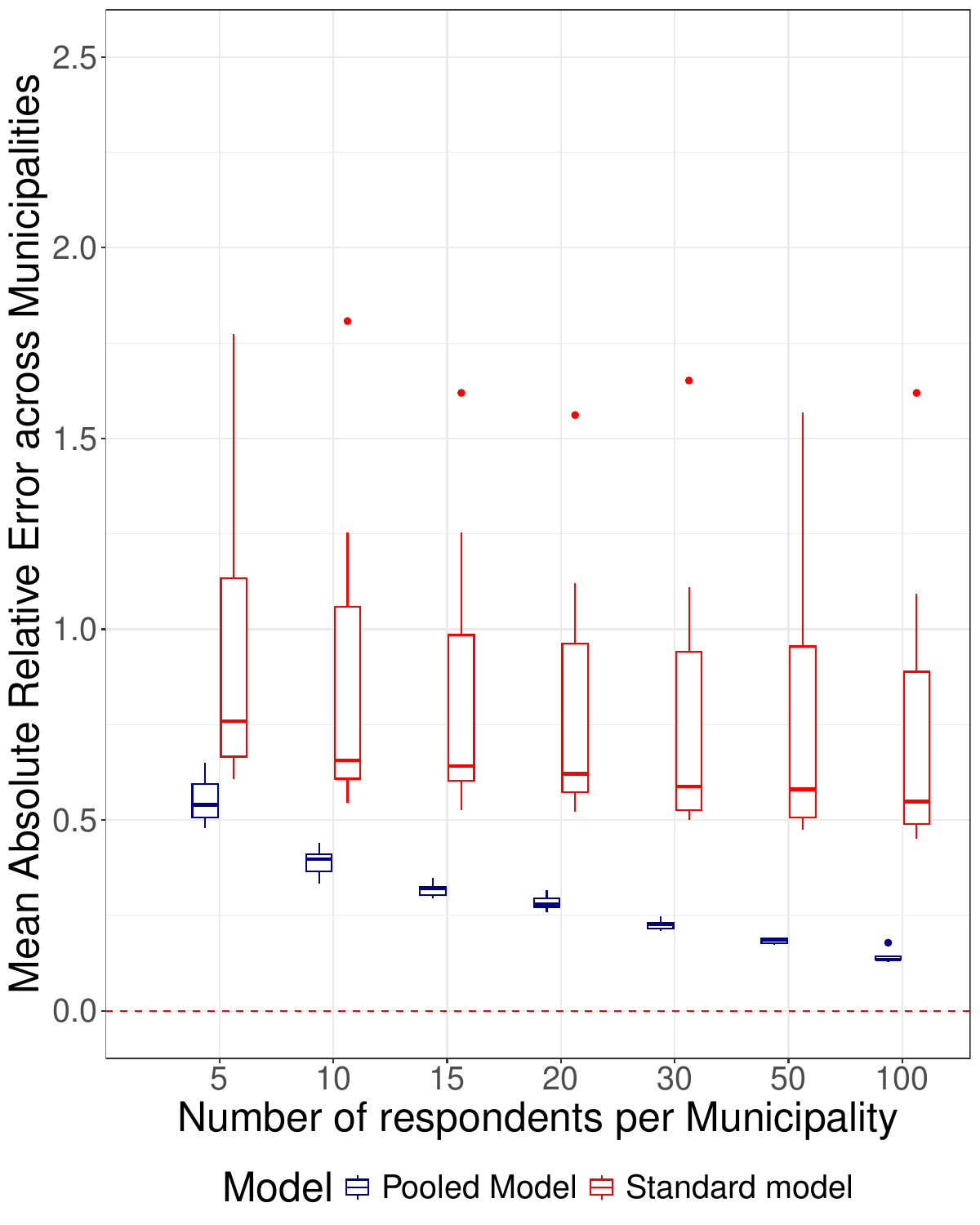}
    \caption{Boxplots showing the MARE. All respondents have a degree of 800 in this data. }
    \label{fig:degree_800}
\end{figure}

To examine the effect of small and large degrees on the error in estimation we simulate two sets of 10 surveys. For one set all the respondents have a degree of 100 and for the second this is 800. The results are shown in Figures \ref{fig:degree_100} and \ref{fig:degree_800}.

\section{Diagnostics for the Philippines Data Set}
\subsection{Trace Plots}
\begin{figure}
    \centering
    \includegraphics[scale=0.6]{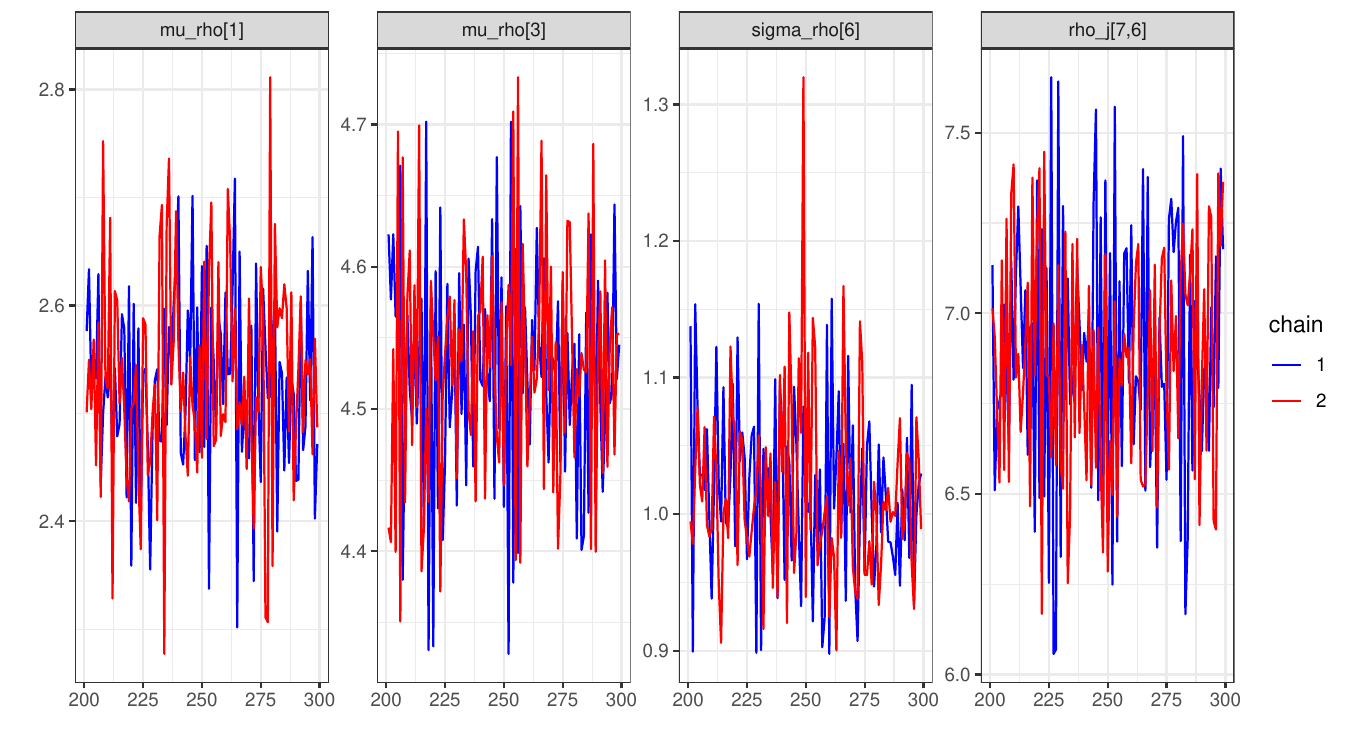}
    \caption{The traceplots for key parameters in the model in one run of the simulation study after warmup iterations are discarded.}
    \label{fig:traceplot_key_parameters}
\end{figure}

Figure \ref{fig:traceplot_key_parameters} shows the convergence of the samples in this simulation study. For quick reference $\mu_{\rho_{1}}  = 2.5$, $\mu_{\rho_{3}} = 4.5$, $\sigma_{\rho_{6}} = 1$ and $\rho_{6,7} = 6.68$. In simulation k and m where transposed for efficiency. 

\FloatBarrier
\newpage

\section{Scaling Estimates for the Philippines Data Set}

\cite{Habecker_2015} introduce the concept of a relative scaling method for NSUM models. Given the estimates and true group sizes for all but one group $k$, a correction term for the estimate for the size of group $k$ in municipality $m$ can be defined as
\begin{equation*}
c_{km} = \frac{1}{a_{km}} \sum^{a_{km}}_{j \in \{{\text{known groups}\}} \setminus k} \frac{\tilde{N}_{jm}} {N_{jm}},
\end{equation*}
where $a_{km}$ is the number of available known groups for group $k$ in $m$, $\tilde{N}_{jm}$ is the estimated size of group $j$ in $m$ and $N_{jm}$ is the true size of group $j$ in $m$. The correction term scales by the average relative errors so that the effects of large and small groups are equal. The idea in this scaling procedure has also been used in NSUM to select known groups in a cross validation method \citep{Habecker_2015}. 

In the Bayesian setting, we follow the approach of \cite{Laga_2023} and scale the posterior estimates for the unknown group sizes using this term. Computationally this can be done by subtracting the log of the correction term from the posterior sample of $\rho_{km}$ for each iteration $i$ or $\rho_{ikm}$. Specifically,
\begin{equation*}
C_{ikm} = \log \left( \frac{1}{a_{km}}  \sum_{j \in \{{\text{known groups}\}} \setminus k} \frac{\exp(\rho_{ijm})}{N_{jm} / N_m}\right),
\end{equation*}
where $N_{m}$ is the total population of $m$. The exponential of the mean of these scaled values for $\rho_{km}$ is then taken as the final estimate for the size of the group in the municipality. Previous approaches used the known group sizes to estimate respondent degrees which then inform the unknown population size estimates or in a Bayesian framework used known group sizes as starting points during sampling for their prevalence parameter as in \citet{Maltiel_2015}. The difference in our approach is that the known group sizes are used after the estimates from the ARD have been established. A Bayesian model that incorporates the known group sizes before may propagate any bias in the known sizes to the estimation of the unknown groups. Additionally the known group sizes can be compared against their estimated values just from the ARD to examine the accuracy of the model. 

\begin{figure}[!ht]
    \centering
    \includegraphics[scale=0.75]{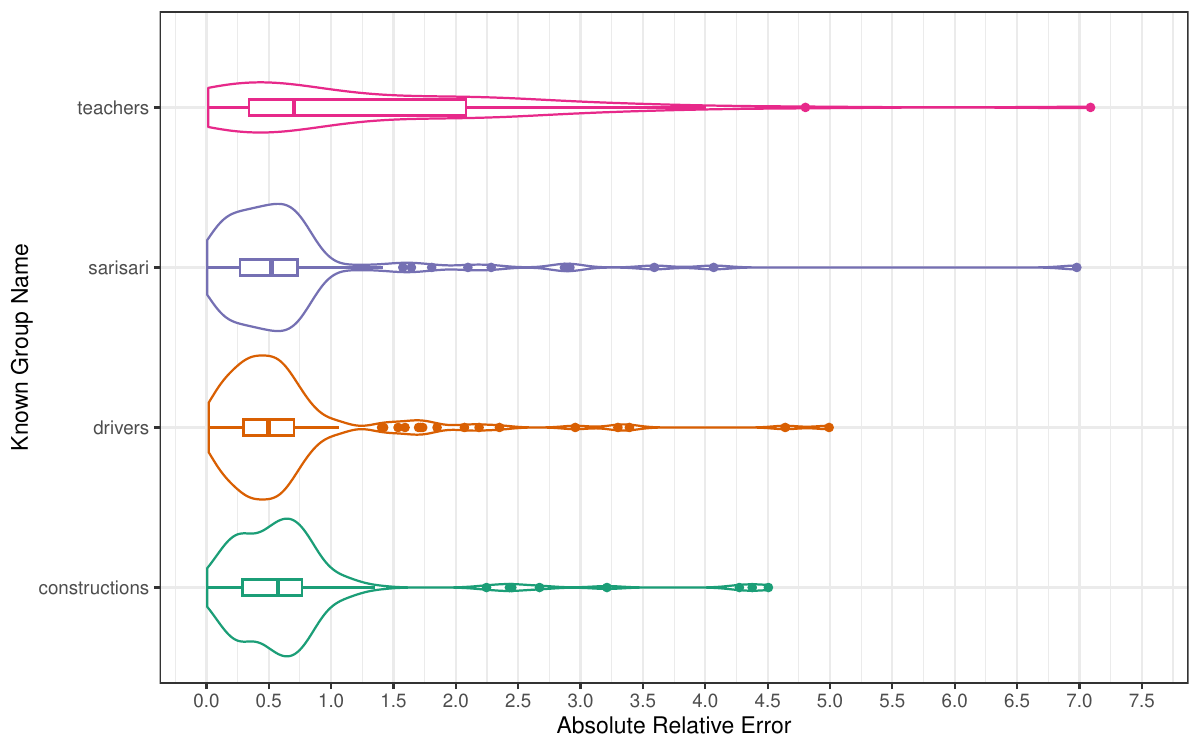}
    \caption{The absolute relative errors of the estimates of each known group for the sampled regions in the Philippines.}
    \label{fig:rel_errors_known_groups}
\end{figure}

To scale the posterior samples generated by our method, the sizes of known groups at the municipality level are needed. We adapted an approach from \citet{Feehan_2022} to estimate the sizes of these groups using demographic data collected in the survey. This allowed us to estimate the number of people within the municipality that belong to the known groups needed. We also collected ARD on these groups from respondents. We validated the accuracy of our approach to the known group sizes by comparing the groups sizes estimated from demographic data to the groups sizes estimated from our Negative Binomial model to the ARD  once scaled. For each known group the estimate from the households was not used to scale our posterior estimates so we can reasonably compare the two approaches for agreement. Figure \ref{fig:rel_errors_known_groups} shows the absolute relative difference in between the two estimates (with the estimates from the households being used as the baseline) across all municipalities. In the majority of municipalities there is a reasonably small relative difference between the two estimates.

\section{Victim Map}
\begin{figure}
    \centering
    \includegraphics[scale=0.65]{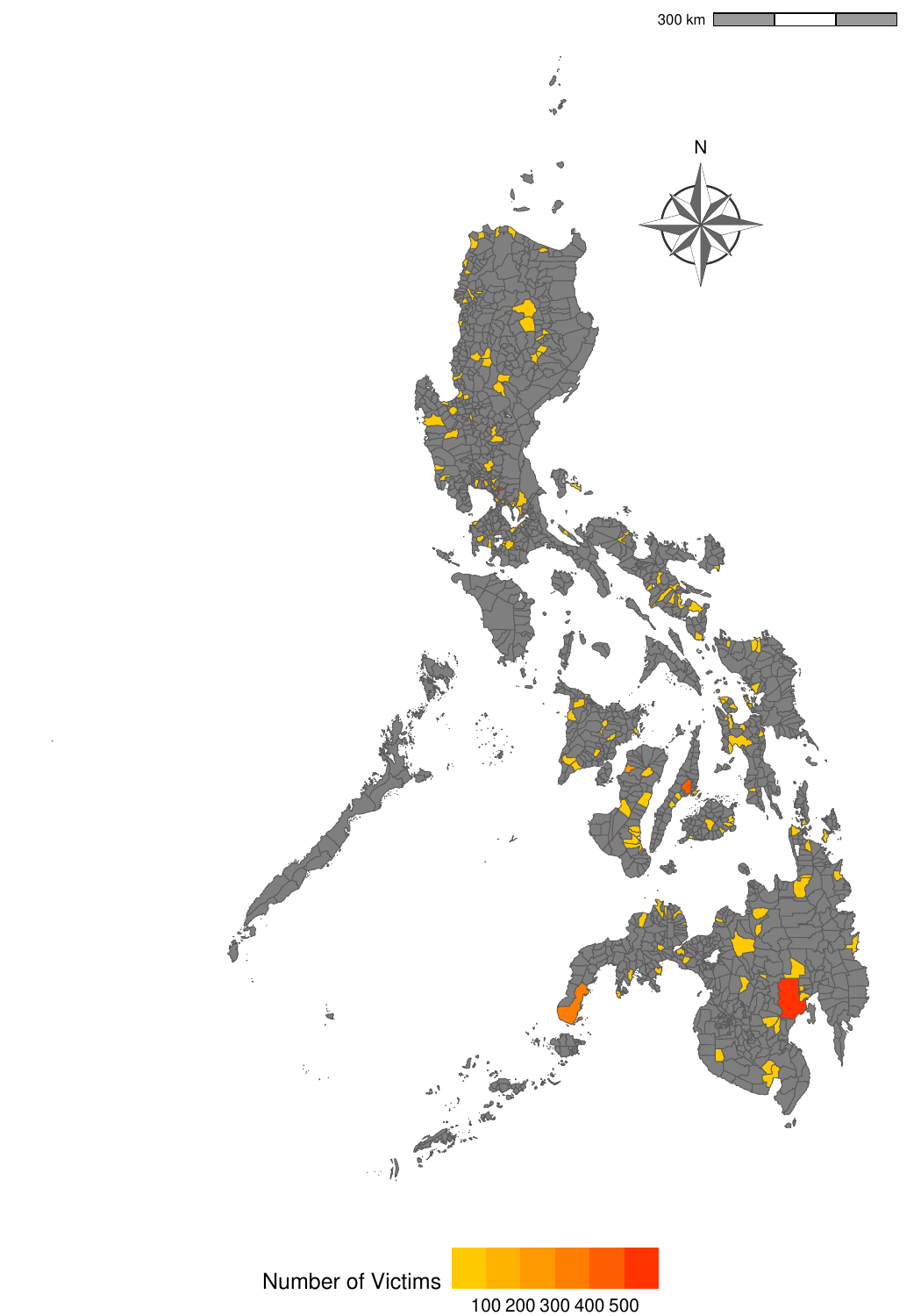}
    \caption{The estimated number of children trafficked by adults to produce child sexual exploitation material across the 150 sampled municipalities.}
\end{figure}

\clearpage
\bibliographystyle{rss}
\bibliography{nsum.bib}